\documentclass[fleqn,10pt]{wlscirep}
\usepackage[utf8]{inputenc}
\usepackage{hyperref}
\usepackage{multirow}
\usepackage{graphicx}
\usepackage{pdfpages}
\usepackage{caption}
\usepackage{amsmath}
\usepackage{amssymb}
\usepackage{booktabs}
\usepackage{xcolor}
\usepackage{colortbl}
\usepackage{bbding}    
\usepackage{pifont}
\definecolor{tablehead}{RGB}{222,220,204}
\definecolor{tablesection}{RGB}{238,236,224}
\definecolor{tablesubtotal}{RGB}{247,246,240}

\newcommand{\pmsp}[1]{$\pm$\small{#1}}
\DeclareMathAlphabet{\cmcal}{OMS}{cmsy}{m}{n}
\newcommand{\Loss}{\cmcal{L}}
\newcommand{\AnatMap}{\cmcal{A}}
\newcommand{\DosePriorMap}{\hat{\cmcal{D}}}
\newcommand{\RiskMap}{\cmcal{R}}
\newcommand{\SpatialPriorMap}{\cmcal{P}}
\newcommand{\safeincludegraphics}[2][]{%
  \IfFileExists{#2}{%
    \includegraphics[#1]{#2}%
  }{%
    \IfFileExists{../#2}{%
      \includegraphics[#1]{../#2}%
    }{%
      \fbox{%
        \begin{minipage}[c][0.32\textheight][c]{0.92\linewidth}
          \centering
          Missing figure file:\\[0.5em]
          \texttt{\detokenize{#2}}
        \end{minipage}%
      }%
    }%
  }%
}
\renewcommand{\thesection}{\Roman{section}}
\renewcommand{\thesubsection}{\thesection.\arabic{subsection}}

\title{Personalising Deformable Image Registration with Clinical Context and Deformation History for Adaptive Radiotherapy}

\author[1]{Caiwen Jiang}
\author[1]{Songlin Zhao}
\author[1]{Yuzhen Ding}
\author[3]{Jia Mi}
\author[1]{Samir H. Patel}
\author[1]{Alexandra Bennett}
\author[1]{Keldon K. Lin}
\author[1]{Terence T. Sio}
\author[1]{Jonathan B. Ashman}
\author[1]{Lisa A. McGee}
\author[1]{Jean-Claude M. Rwigema}
\author[1]{William G. Rule}
\author[1]{Sameer R. Keole}
\author[1]{Sujay A. Vora}
\author[1]{William W. Wong}
\author[1]{Nathan Y. Yu}
\author[1]{Michele Y. Halyard}
\author[1]{Steven E. Schild}
\author[1]{Safia K. Ahmed}
\author[2]{Daniel J. Ma}
\author[2]{Robert L. Foote}
\author[2]{Michelle A. Neben Wittich}
\author[4]{Thomas J. Whitaker}
\author[5]{Haibo Lin}
\author[3]{Dinggang Shen}
\author[1,*]{Wei Liu}

\affil[1]{Department of Radiation Oncology, Mayo Clinic, Phoenix, Arizona, USA}
\affil[2]{Department of Radiation Oncology, Mayo Clinic, Rochester, Minnesota, USA}
\affil[3]{School of Biomedical Engineering, ShanghaiTech University, Shanghai, China}
\affil[4]{Department of Radiation Physics, The University of Texas MD Anderson Cancer Center, Houston, Texas, USA}
\affil[5]{New York Proton Center, New York, New York, USA}
\affil[*]{Correspondence to Wei Liu (liu.wei@mayo.edu).}

\begin{abstract}
Adaptive radiotherapy must account for anatomical changes during treatment, but
doing so depends on reliably mapping targets, organs at risk and dose across
serial images. Existing registration methods treat each image pair independently,
without using radiotherapy-specific context or the patient's preceding anatomical
trajectory. We developed Clinical-context-Aware and Patient-adaptive Deformable
Image Registration (CAP-DIR), integrating source contours, dose, clinical text
and prior-registration features. We evaluated 16,361 CT examinations from 3,473
patients across four institutional cohorts and five public benchmarks. CAP-DIR
increased internal-test organ-at-risk-region normalized cross-correlation from
84.76\% to 96.92\% relative to the strongest learning-based comparator, with low
deformation folding, and remained superior across seven external cohorts. In a
blinded review of 20 longitudinal cases, clinically acceptable contours increased
from 45.0\% to 82.5\% for clinical target volumes and from 22.5\% to 62.5\% for
organs at risk, while estimated editing time decreased. In a controlled benchmark,
CAP-DIR reduced global dose-mapping error by 34.5\% relative to the strongest
comparator. These findings show that clinically informed, longitudinal
registration can improve contour usability and dose mapping for adaptive
radiotherapy.
\end{abstract}

\begin{document}
\maketitle

Adaptive radiotherapy aims to maintain treatment accuracy as patient anatomy
changes over a course of repeated treatment fractions. Tumour regression, organ
filling, weight loss and other anatomical changes can alter the relationship
between the planned and delivered dose
\cite{sonke2019adaptive,glidehurst2021adaptive,hall2022magnetic}. This treatment
strategy responds to these changes by acquiring treatment-state images and
modifying the plan when anatomical or dosimetric deviations become clinically
important \cite{yan1997adaptive,lavrova2023adaptive}. Deformable image
registration (DIR) provides the spatial correspondence needed to propagate
contours and map dose between the planning image and subsequent images. Its
accuracy therefore directly affects the information available for adaptive
assessment. Reliable and efficient longitudinal mapping is therefore not merely
a technical component of adaptive radiotherapy but a prerequisite for translating
repeated treatment imaging into clinically actionable adaptation.
These demands differ from those of registration in a single diagnostic imaging
study. A radiotherapy course is organized around a treatment plan, defined
targets, organs at risk (OARs) and spatially heterogeneous dose objectives. Small
misregistrations can be inconsequential in homogeneous soft tissue yet important
near a target boundary, serial organ or steep dose gradient. The consequences
also extend beyond a numerical registration score. Inaccurate propagation can
increase contour-editing workload, distort structure-level dose estimates and
delay decisions about whether replanning is needed. A clinically useful DIR
system should therefore preserve broad anatomical correspondence while assigning
particular importance to the structures and regions that determine treatment.

Most DIR algorithms nevertheless formulate each registration as a generic and
independent image-matching problem. Classical optimization methods can be slow in
time-sensitive workflows, whereas learning-based methods such as
VoxelMorph~\cite{balakrishnan2019voxelmorph},
TransMorph~\cite{chen2022transmorph}, CorrMLP~\cite{wang2024corrmlp},
MambaMorph~\cite{wang2024mambamorph},
DiffuseMorph~\cite{kim2022diffusemorph} and
UniGradICON~\cite{tian2024unigradicon} primarily learn population-level image
correspondence. Global image similarity may underrepresent smaller or
low-contrast organs at risk (OARs), although errors in these regions can alter
contour review and dose assessment. Pairwise methods also discard the sequence
of anatomical states already observed for the same patient.
This independence assumption is especially restrictive in fractionated
radiotherapy. The current image is not an unrelated sample. It follows earlier
images acquired under the same treatment course, and those images may reveal
patient-specific patterns such as reproducible organ filling, progressive weight
loss or recurring respiratory displacement. Population-level training captures
common deformation patterns, but it does not explicitly use the trajectory
already observed for an individual patient. Conversely, patient-specific
optimization at inference may be too slow or unstable for routine online use.
Representing preceding registrations as a fixed-length memory offers a way to
use longitudinal evidence without retraining the model for every patient.

The radiotherapy record provides two complementary sources of information that
could address these limitations. The first is multimodal clinical context.
Structure contours identify targets and OARs for which propagation errors require
clinical review, dose distributions
identify regions where small spatial errors can alter dose estimates, and
clinical descriptions provide semantic information about the treatment site,
diagnosis and plan. Together, these spatial and semantic signals define where
registration accuracy is most consequential. The second is longitudinal
deformation history. Earlier treatment images and registration outputs document
the patient's evolving anatomical trajectory,
including recurring motion and progressive treatment-related change. Recent
advances in multimodal learning~\cite{perez2018film,singhal2023large,moor2023foundation} and
sequential modelling~\cite{cho2014learning} make it possible to use these
complementary resources in a unified registration system.

We therefore developed Clinical-context-Aware and Patient-adaptive Deformable
Image Registration (CAP-DIR), which integrates source-side clinical context with
a temporally ordered memory of previous registrations. Clinical context directs
registration towards treatment-relevant anatomy, while longitudinal memory adapts
the inference to the anatomical trajectory already observed for the individual
patient. To evaluate CAP-DIR, we used 16,361 CT examinations from 3,473 patients
across four institutional
adaptive-radiotherapy cohorts and five public longitudinal imaging benchmarks,
spanning seven anatomical sites and more than 20 disease categories
(Fig.~\ref{fig:overview}). We first assessed image similarity and deformation
regularity on a patient-level internal test set and seven held-out external
cohorts. We then examined anatomical and clinical relevance using longitudinal
CT pairs with manual contours, a blinded clinician review of propagated contours
and a controlled dose-mapping benchmark. Separate analyses quantified the
contributions of clinical context and preceding patient history. All conditioning
inputs were restricted to source-side or preceding information, whereas
target-side contours were used only as evaluation references. This design tested
whether CAP-DIR
improves not only image similarity but also the anatomical correspondence and
downstream tasks that support adaptive radiotherapy.

\section{Results}

\subsection{Study design and cohorts}

CAP-DIR estimates a source-to-target deformation field from a planning or
reference CT and a subsequent treatment-state CT. Stage~1 conditions the
registration backbone on source-side contours, dose and structured clinical
text. Stage~2 additionally summarizes image features, deformation statistics
and quality measures from earlier registrations for the same patient
(Fig.~\ref{fig:overview}a,c). Target-side contours were withheld from the
model and used only for evaluation.
The source-side information reflects what would be available before registering
the current image. Structure contours define treatment-relevant anatomy, the dose
distribution provides a spatial representation of treatment emphasis, and the
clinical text describes variables such as disease site, diagnosis and treatment
context. In the longitudinal stage, the memory contains only registrations that
precede the current target image. The model therefore does not use future images
or target-side contours. Training and model selection were performed separately
for the two stages using the internal validation partition, and the selected
models were then fixed for testing.

The dataset comprised 3,473 patients, 5,005 plans, 16,361 CT examinations and
11,986 source--target image pairs (Fig.~\ref{fig:overview}b and
Table~\ref{tab:dataset_split}). Model development used patient-level partitions
of the Mayo Clinic Arizona and Minnesota cohorts. These included 2,052 patients
and 7,429 pairs
for training, 293 patients and 1,062 pairs for validation, and 586 patients and
2,122 pairs for internal testing. MD Anderson Cancer Center, the New York Proton
Center and five public benchmarks were held out for external evaluation,
comprising 542 patients and 1,373 pairs. Pair construction and the derivation of
clinical-context inputs for public datasets are described in the Methods.
The cohorts covered seven anatomical sites and more than 20 disease categories,
with different scanners, imaging protocols and longitudinal structures. In
institutional cohorts with fraction-level verification imaging, a planning CT
was paired with eligible subsequent verification CTs from the same plan. When a
course contained multiple planning CTs but no fraction-level verification CT,
the first planning CT was paired with eligible later planning CTs. For public
longitudinal benchmarks, the baseline or reference CT was paired with eligible
later examinations from the same patient or imaging sequence. These fixed pairs,
rather than randomly redefined pairs, were used for evaluation.

\subsection{Internal and external registration performance}

On the internal test set, CAP-DIR with clinical context and patient history
achieved the highest image-similarity performance, with body-region normalized
cross-correlation (NCC) of 98.61\%, OAR-region NCC of 96.92\%, body-region
structural similarity of 94.42\% and OAR-region structural similarity of 93.36\%
(Fig.~\ref{fig:registration_performance}a). CorrMLP was the strongest
general-purpose learning-based comparator, reaching body-region NCC and
structural similarity of 96.08\% and 89.94\%, and OAR-region values of 84.76\%
and 88.93\%, respectively. CAP-DIR therefore improved the two body-region
measures by 2.53 and 4.48 percentage points and the two OAR-region measures by
12.16 and 4.43 percentage points. The clinical-context model without patient
history already reached OAR-region NCC of 94.88\% and structural similarity of
90.74\%, with longitudinal conditioning providing the remaining improvement.
CAP-DIR also retained a low OAR folding ratio of 0.08\%, compared with 0.91\%
for CorrMLP, and an OAR-region Jacobian deviation of 0.0038, indicating that the
accuracy gain was not obtained through widespread non-physical deformation.
Detailed internal results are provided in Supplementary Tables~1 and 2.

The internal improvement was concentrated in the clinically specific OAR region.
The body-to-OAR NCC gap ranged from 12.49 to 21.30 percentage points for
TransMorph, UniGradICON, MambaMorph, VoxelMorph and DiffuseMorph and was 11.32
percentage points for CorrMLP. Clinical context reduced this gap to 3.04
percentage points, and adding patient history reduced it further to 1.69
percentage points (Fig.~\ref{fig:registration_performance}c). This regional
advantage was observed across all seven anatomical sites
(Fig.~\ref{fig:registration_performance}b). Relative to CorrMLP, OAR-region NCC
increased by 6.0 percentage points in lung, 19.6 in pelvis, 17.8 in abdomen, 13.8
in brain and 11.1 in head-and-neck cases, whereas body-region gains ranged from
1.0 to 4.3 percentage points. The overall improvement was therefore neither
limited to broad body alignment nor driven by a single anatomical cohort.

Without retraining on the held-out data, CAP-DIR achieved the highest values for
all four reported image-similarity measures across the seven external cohorts
(Fig.~\ref{fig:registration_performance}d,e). Absolute OAR-region NCC was 95.8\%
in NLST, 89.5\% in Longitudinal CT, 97.8\% in RIDER, 82.5\% in Pancreas-CT,
99.5\% in 4D-Lung, 92.8\% in the New York Proton Center cohort and 92.2\% in the
MD Anderson cohort. Gains over the strongest comparator within each cohort ranged
from 0.8 to 14.3 percentage points for OAR-region NCC and from 1.2 to 11.8
percentage points for structural similarity, including OAR-region NCC gains of
6.2 and 6.3 percentage points in the New York Proton Center and MD Anderson
cohorts, respectively. The greatest improvement occurred in Pancreas-CT, whereas
the smaller gain in 4D-Lung reflected near-ceiling baseline performance.
Although absolute performance varied with anatomy and image source, the direction
of the comparison was consistent across public and institutional datasets.
Cohort-level numerical results are provided in Supplementary Table~3.

\subsection{Contour propagation and blinded review}

Because improved image similarity alone does not establish clinical usefulness,
we next tested whether the predicted correspondence produced more accurate and
clinically usable contour propagation. This analysis used longitudinal planning
CTs with clinician-delineated contours on
both the source- and target-side images, including 276 first-to-subsequent pairs
from the internal cohort and 44 additional pairs from the two private external
cohorts. CAP-DIR increased OAR Dice similarity coefficient (DSC) to 92.07\%,
compared with 82.28\% for CorrMLP, and increased clinical target volume (CTV) DSC
to 87.14\%, compared with 73.32\% (Fig.~\ref{fig:contour}a--c). Across structures
and sites, this represented an overall DSC improvement of 57\% relative to
unregistered contour transfer, 30\% relative to Demons~\cite{thirion1998image},
a conventional intensity-based algorithm commonly used in clinical registration
systems, and 14\% relative to CorrMLP. The residual symmetric-difference volume
between propagated contours
and target-side reference contours decreased from 9.66 to 2.06~ml on the common
1-mm grid (Fig.~\ref{fig:contour}d,e), corresponding to reductions of 79\%, 70\%
and 55\% relative to direct transfer, Demons and CorrMLP, respectively.
Site-specific reductions relative to direct transfer ranged from 74\% to 81\%.
These findings provided anatomical evidence independent of the intensity metrics
used for model development. Additional cohort and residual-volume details are
provided in the Supplementary Note ``Contour propagation and residual contour
volume''.

We next examined whether the quantitative improvements translated into clinically
usable propagated contours. Two radiation oncology reviewers performed a
method-blinded assessment of 20 longitudinal cases, with four cases each from the
abdomen, brain, breast or chest wall, head and neck, and pelvis. Each case
contained the clinician-approved reference contours and two unedited contour sets
propagated by CAP-DIR and CorrMLP. The reference set was identified, whereas the
two propagated sets were shown on the same target-side image using randomized
anonymous labels. The review was performed in the clinical ARIA environment at
Mayo Clinic Arizona using the image display and contour-review interface used in
routine radiotherapy practice. Reviewers rated CTVs and OARs separately on a four-point scale.
A score of 4 indicated clinical acceptability without modification, 3 indicated
the need for minor edits, 2 indicated the need for major edits, and 1 indicated
that the contours were unusable. Scores of 3 or 4 were classified as clinically
acceptable. Reviewers also estimated the time required to make each result
clinically usable. The proportion rated clinically acceptable increased from
45.0\% with CorrMLP to 82.5\% with CAP-DIR for CTVs and from 22.5\% to 62.5\%
for OARs
(Fig.~\ref{fig:contour}f). Median clinician-estimated editing time decreased from
11.3 to 8.8 min for CTVs and from 26.3 to 15.0 min for OARs
(Fig.~\ref{fig:contour}g and Supplementary Table~4). Patient-level bootstrap
confidence intervals and inter-reviewer agreement are reported in Supplementary
Table~4.

The largest absolute increase in clinical acceptability occurred for OARs, which
often comprise several propagated structures and therefore create a greater
editing burden. The agreement between the objective contour analysis and the
reviewers' assessments indicates that the anatomical improvement extended to
clinical usability.

\subsection{Controlled dose-mapping benchmark}

We next assessed whether the registration improvement reduced error in
accumulated-dose mapping. Because voxel-wise ground-truth accumulated dose is
unavailable in clinical data, we used a controlled synthetic benchmark that
generated a reference dose in the planning state and treatment-state doses under
prescribed treatment-course deformations (Fig.~\ref{fig:dose}a). The known
deformations allowed the recovered accumulated dose to be compared voxel by voxel
with ground truth. Direct transfer produced a mean absolute error of 1.83\% of the
dose scale, compared with 1.43\% for Demons, 1.10\% for CorrMLP and 0.72\% for
CAP-DIR (Fig.~\ref{fig:dose}b). CAP-DIR therefore reduced global dose error by
34.5\% relative to CorrMLP.

At the structure level, CAP-DIR reduced the average error across OAR and CTV
mean- and maximum-dose endpoints from 0.97\% with CorrMLP to 0.64\%
(Fig.~\ref{fig:dose}c). CTV mean-dose error decreased from 1.10\% to 0.76\%, CTV
maximum-dose error from 0.97\% to 0.63\%, OAR mean-dose error from 0.87\% to
0.58\%, and OAR maximum-dose error from 0.92\% to 0.60\%. Reductions were also
observed for regional, high-gradient and dose--volume-histogram errors across
anatomical sites (Fig.~\ref{fig:dose}d,e). These findings isolate the effect of
registration on dose mapping under known spatial transformations. The reductions
in structure-level and high-gradient errors link improved spatial correspondence
to the reliability of dose quantities used in adaptive assessment. They do not
constitute prospective validation of clinical delivered-dose accumulation or
capture all sources of uncertainty in that process. Additional site-specific
results are provided in the Supplementary Note ``Controlled dose-mapping
results'' and Supplementary Table~5.

\subsection{Contribution of clinical context}

We compared image-only registration with models conditioned on structured
clinical text, spatial context from contours and dose, or both
(Fig.~\ref{fig:clinical_context_effect}a). All Stage~1 ablations used the same
architecture, data partition, training schedule and OAR-weighted objective, so
they differed only in the context supplied to the model. Image-only registration
reached a body-region NCC of 95.8\% but an OAR-region NCC of 84.4\%. Structured
clinical text increased OAR-region NCC to 90.1\%, spatial context increased it to
92.4\%, and their combination increased it to 94.9\%
(Fig.~\ref{fig:clinical_context_effect}b). The combined model also increased
body-region NCC to 97.9\%, producing the highest performance in both regions.

The distribution of the gains showed that clinical context contributed more in
OARs than in the body region. Text conditioning increased body- and OAR-region
NCC by 1.6 and 5.7 percentage points, respectively, whereas contours and dose
increased them by 1.2 and 8.0 percentage points. Combining both sources increased
body-region NCC by 2.1 percentage points and OAR-region NCC by 10.5 percentage
points (Fig.~\ref{fig:clinical_context_effect}c,d), narrowing the body-to-OAR gap
from 11.4 to 3.0 percentage points. This regional benefit was observed across all
anatomical sites. The combined context improved OAR-region NCC by 5.4 percentage
points in lung, 6.3 in breast, 14.9 in abdomen and 16.5 in pelvis, with the
largest gains in abdominal and pelvic cases
(Fig.~\ref{fig:clinical_context_effect}e).
Thus, clinical context did not simply increase global image similarity but
redirected registration gains towards treatment-relevant anatomy.

\subsection{Contribution of patient history}

The longitudinal memory combined features from earlier images with summaries of
the predicted deformation and its registration quality. Stage~2 used only
observations preceding the current target-side image and did not perform
patient-specific fine-tuning at inference. We compared this history-conditioned
model with the clinical-context-only model across the available history-length
strata (Fig.~\ref{fig:adaptive}). With clinical context alone, OAR-region NCC
ranged from 95.5\% among cases with one previous observation to 92.5\% among
those with 17. The history-conditioned model ranged from 97.1\% to 96.0\% across
the same strata and improved OAR-region NCC at every evaluated history length.

The history-related OAR gain ranged from 1.6 percentage points among cases with
one previous observation to 3.5 percentage points among cases with 17, whereas
the body-region gain ranged from 0.6 to 1.0 percentage points. Gains were
0.9--2.0 percentage points in brain and breast cases and 2.4--2.7 percentage
points at shorter histories in head-and-neck and pelvic cases. Abdominal cases
showed the largest site-specific benefit, ranging from 2.2 to 5.2 percentage
points across the reported strata. Two patterns emerged from these results. The
larger gains in OAR than in body regions indicate that longitudinal memory
primarily refined structure-focused correspondence rather than uniformly
increasing global alignment. Across the observed strata, the history-conditioned
model maintained OAR-region NCC between 96.0\% and 97.1\%, whereas the
clinical-context-only model ranged from 92.5\% to 95.5\%. The larger benefits in
abdominal, head-and-neck and pelvic cases are also consistent with greater value
from patient-specific history in settings characterized by recurring organ motion
or progressive anatomical change. Together, these findings show that preceding
registration information complemented population-level clinical context and
supported more stable alignment of treatment-relevant anatomy across longitudinal
imaging. The observed history-length ranges are summarized in the Supplementary
Note ``History-length stratification''.

\section{Discussion}

In a large multi-cohort evaluation, CAP-DIR frames deformable registration in
adaptive radiotherapy as a clinically conditioned and longitudinal inference
problem. Across internal and external cohorts, its largest improvement occurred
in OAR regions rather than in broad body regions. This regional advantage was
maintained across seven external cohorts and extended beyond image-similarity
measures. Evaluation against
clinician-approved contours showed more accurate anatomical correspondence, the
blinded review found a higher proportion of clinically acceptable propagated
contours with less editing, and the controlled dose-mapping benchmark showed
lower errors in accumulated-dose maps and structure-level dose summaries.
The concordance between regional registration metrics, contour propagation,
blinded clinical review and controlled dose mapping indicates that the
improvement extended beyond image similarity to tasks directly involved in
adaptive assessment.

The concentration of these gains in OAR regions is particularly relevant because
whole-image similarity can be dominated by large, relatively stable regions. A model may
therefore achieve a high global score while remaining inaccurate at smaller
structures or interfaces where propagation errors require clinical review.
CAP-DIR reduced the body-to-OAR performance gap by using radiotherapy planning
information to identify where registration accuracy was most consequential. The
agreement between regional image metrics, manual-contour evaluation and blinded
review indicates that this conditioning changed the anatomical distribution of
registration accuracy rather than merely increasing global similarity. In the
clinical review, the improvement was observed for both CTVs and OARs and was
particularly marked for OAR acceptability and editing requirements. These
results support the use of CAP-DIR to provide a more accurate starting point for
clinician review rather than to replace that review.

The ablation analyses provide insight into the source of these clinically
focused gains. Source
contours and dose maps supplied spatial information about treatment-relevant
anatomy, while structured clinical text supplied semantic information about the
treatment site, diagnosis and plan. Combining these forms of clinical context
produced larger OAR gains than either source alone. The temporally ordered memory
then added patient-specific information from preceding images, predicted
deformations and registration-quality features, improving performance beyond the
clinical-context-only model across the evaluated history strata and anatomical
sites. These two components also have complementary practical roles. Planning
contours, dose and structured metadata can be reused from treatment preparation,
whereas the memory is updated as treatment images become available. CAP-DIR can
therefore adapt its inference to the observed anatomical course while retaining
a fixed population-trained model and requiring neither target-side contours nor
patient-specific optimization at inference.

Beyond identifying the contributions of clinical context and patient history,
the external evaluation and dose benchmark tested whether the resulting gains
generalized across imaging settings and extended to a downstream dosimetric
task. Independent radiotherapy cohorts and public longitudinal benchmarks
varied in anatomy, acquisition and follow-up, producing variation in absolute
performance, yet CAP-DIR maintained its relative advantage. This consistency
supports transportability across imaging settings, although performance remains
most appropriately interpreted within each cohort. The dose benchmark addressed
a distinct consequence of spatial correspondence. Because a true patient
deformation is unavailable for voxelwise validation, prescribed deformations
provided a controlled reference for measuring dose-mapping error. Reductions in
voxel-level, high-gradient and structure-level errors showed that the gains in
registration propagated to dosimetric quantities. Taken together with the
contour review, these experiments position the method as a form of clinical
assistance in which propagated contours and mapped dose support assessment and
adaptation while remaining subject to clinician review and quality control.

These findings should nevertheless be interpreted in light of several
limitations. Although CAP-DIR showed consistent performance across seven
held-out external cohorts, model development was based on data from two centres
within the Mayo Clinic system, and broader prospective evaluation across
additional institutions and practice settings is warranted. The scope of
generalisability is further bounded by the present focus on longitudinal
CT-to-CT registration, leaving performance with other imaging modalities and
cross-modality registration to be established. Within this evaluated setting,
clinician-approved contours provided a clinically relevant anatomical reference,
but they remain subject to observer variability rather than representing an
absolute anatomical ground truth. Beyond these constraints on generalisability
and the reference standard, the longitudinal design introduces a model-specific
dependency because its memory incorporates deformation and quality features from
preceding Stage~1 registrations. Errors in these earlier estimates could
therefore propagate into subsequent registrations, and the robustness of the
memory mechanism to such error propagation remains to be evaluated.

\section{Conclusion}

CAP-DIR demonstrates that deformable image registration for adaptive radiotherapy
can benefit from integrating radiotherapy-specific clinical context with
longitudinal patient history. Across internal testing and seven held-out external
cohorts, this formulation preferentially improved registration in
treatment-relevant OAR regions while maintaining plausible deformations. The
gains extended to contour propagation, blinded clinical assessment and
controlled dose mapping, linking improved anatomical correspondence with tasks
central to adaptive radiotherapy. These findings support clinically informed and
patient-adaptive registration as a means of providing more reliable starting
contours and longitudinal dose estimates for adaptive assessment across the
treatment course.

\section{Methods}

\subsection{Ethics}
This multi-site retrospective study was approved by the Mayo Clinic
Institutional Review Board (IRB \#: 24-011106), which served as the single IRB
of Record for the participating sites, and was conducted in accordance with the
Declaration of Helsinki. Given the retrospective design, the Mayo Clinic
Institutional Review Board waived the requirement for study-specific written
informed consent. Data transfer and analysis were performed under the
corresponding institutional data-use arrangements. Publicly available
de-identified datasets were used in accordance with their original access and
data-use requirements.

\subsection{Study datasets and preprocessing}
\label{sec:data}

\noindent\textbf{Longitudinal radiotherapy cohorts.}
Institutional longitudinal radiotherapy data were collected from Mayo Clinic
Arizona, Mayo Clinic Minnesota, MD~Anderson Cancer Center and the New York Proton
Center, and public longitudinal CT benchmarks were included to broaden the
imaging protocols and anatomical settings represented in evaluation. Across
cohorts, available information included CT images acquired across the treatment
course, structure contours, dose distributions and patient-level clinical text.
For public datasets that did not contain complete clinical information across
the treatment course, the clinical-context inputs required by the model were
completed using a standardised preprocessing pipeline. Structured clinical text
was assembled from available de-identified dataset annotations, DICOM header
fields and imaging metadata. Source structure masks were obtained from available
baseline or reference annotations and supplemented, when necessary, using the
fixed pretrained TotalSegmentator model~\cite{wasserthal2023totalsegmentator}.
When clinical dose distributions were unavailable, source-space proxy dose maps
were generated from the available source target contours using a standardised
geometric distance-to-target formulation~\cite{appenzoller2012predicting}. All
derived clinical-context inputs were constructed from source-side information,
while target annotations were reserved exclusively for evaluation.
All records were organised by patient, treatment course and treatment plan to
preserve the longitudinal relationship among plans, fractions and adaptive
imaging time points. Detailed cohort composition, patient-level partitioning and
image-pair counts are provided in Table~\ref{tab:dataset_split}.

\noindent\textbf{Preprocessing.}
Preprocessing was performed separately for imaging data and clinical text. For
CT examinations from the same patient, planning or reference CTs and
verification or treatment-state CTs were first reoriented to a common
anatomical orientation and harmonised in physical space to account for
differences in image origin, direction, matrix size, voxel spacing and field of
view. CT volumes were then resampled to an isotropic resolution of
$1\times1\times1$~mm$^3$ and cropped or padded to a fixed matrix size of
$512\times512\times256$. CT intensities were clipped to a fixed HU range of
$[-1000,1000]$ to retain air, soft-tissue and high-density anatomical
information relevant to radiotherapy registration, and were linearly normalised
to $[-1,1]$. Structure masks were transformed using the same spatial operations
with nearest-neighbour interpolation. Dose maps derived from the treatment-planning
dose grid were first mapped to the physical coordinate system of the
corresponding CT and then subjected to the same resampling, cropping and padding
operations as the CT volume using linear interpolation. Dose values were kept in
Gy during preprocessing, and voxels outside the original dose grid were assigned
zero dose.

Clinical text was extracted from CT planning records, treatment-plan information
and clinical notes, and then organised into a structured template before model
training. The template summarised the anatomical site, diagnosis, treatment plan,
prescribed dose and fractionation, image-acquisition information and relevant
structure context, including available OAR and CTV information.

\subsection{Image-pair construction and dataset partitioning}

Throughout the primary registration analyses, a registration case denotes one
source--target CT pair together with the auxiliary inputs available to the
evaluated model. For CAP-DIR Stage~1, these inputs comprise the image pair and
its corresponding clinical context; Stage~2 additionally includes the patient's
available prior treatment history. The source--target pair uniquely defines the
case, so adding clinical context or history does not change the case count.
Image-pair construction depended on the imaging structure available in each
cohort. For cohorts with fraction-level verification imaging, each planning CT
was paired with every eligible subsequent verification CT from the same treatment
plan. For cohorts in which each plan contained only one planning CT and no
fraction-level verification CT, plans were ordered chronologically within each
patient and treatment course, and the planning CT from the first plan was paired
with each eligible planning CT from subsequent plans. For public longitudinal
imaging benchmarks without explicit plan--fraction structure, the baseline or
reference CT was paired with eligible later CT examinations from the same patient
or imaging sequence. The resulting
dataset contained 3,473 patients, 5,005 treatment plans, 16,361 CT examinations
and 11,986 image pairs. Dataset-level counts are reported in
Table~\ref{tab:dataset_split}.

Mayo~AZ and Mayo~MN patients were randomly split at the patient level into
training, validation and internal test sets using a 70\%, 10\% and 20\% split,
respectively. All scans from the same patient were assigned to only one split to
prevent patient-level data leakage. This produced 2,052 training patients with
7,429 image pairs, 293 validation patients with 1,062 image pairs and 586
internal test patients with 2,122 image pairs. MD~Anderson, NYPC and all five
public datasets were held out entirely for external validation, comprising 542
patients and 1,373 image pairs.

Within these fixed patient-level splits, Stage~1 treated the image pairs as
independent registration cases. For Stage~2, samples were organised as a current
image pair together with a temporally ordered sequence of previous image pairs
from the same patient. Only prior fractions or plans were included in the history
for a given current target image, preventing leakage from the current or future
time points. During training, current cases were randomly drawn from the
corresponding fixed split.

\subsection{CAP-DIR framework}
\label{sec:capdir_method}

The core CAP-DIR network estimates a dense source-to-target deformation field
from a source CT $I_s$ and target CT $I_t$ (Fig.~\ref{fig:method_framework}a).
The two CT volumes are processed by weight-shared 3D CNN encoders to extract
hierarchical image features. Each encoder contains five resolution levels, with
two $3\times3\times3$ convolutional layers, instance normalisation and LeakyReLU
activation at each level, followed by downsampling to construct a feature
pyramid. Corresponding source and target features are fused at each resolution,
forming multi-scale representations $\{\mathbf{F}_l\}_{l=1}^{5}$ for
deformation refinement.

The fused feature pyramid is passed to a progressive transformer decoder. From
the coarsest to the finest resolution, each decoder level models long-range
contextual dependencies with a Swin Transformer block and predicts a residual
displacement field $\Delta\phi_l$. The deformation field from the previous
coarser level is upsampled and additively composed with the residual update:
\begin{equation}
  \phi_l = \mathrm{Up}(\phi_{l+1}) + \Delta\phi_l .
\end{equation}
This progressive refinement provides a stable coarse initialisation for large
deformations while preserving finer anatomical detail at higher resolutions. The
final deformation field $\phi_{s\rightarrow t}$ is used to warp the source CT by
trilinear interpolation, enabling image-similarity loss computation and
registration assessment. The multimodal clinically informed stage (Stage~1) and
the patient-adaptive stage (Stage~2) use the same image-registration backbone.
CAP-DIR was therefore formulated as a conditional registration framework.
Stage~1 conditions the shared registration backbone on the current multimodal
clinical context, whereas Stage~2 further conditions the same backbone on
longitudinal patient-history information through a memory encoder.

\noindent\textbf{Clinical-context conditioning.}
All clinical-context inputs, including structure contours, dose maps and planning
text, were restricted to source-side information available before registration.
Manual contours on target CTs, when available, were withheld from the model and
used exclusively for evaluation.
As shown in Fig.~\ref{fig:method_framework}b,c, the clinical-context module used
two complementary sources of information from the radiotherapy workflow.
Semantic context was provided by the structured text prompt, which summarised the
anatomical site, diagnosis, treatment plan, prescribed dose and fractionation,
image-acquisition information and relevant structure context. Spatial context
was provided by structure contours and dose maps, which indicate anatomy and
treatment-sensitive regions where registration errors are more clinically
consequential. These semantic and spatial inputs were converted into decoder
conditioning signals as follows.

As illustrated in Fig.~\ref{fig:method_framework}b, semantic context was first
standardised into a structured text prompt before being passed to the
registration network. This step converted heterogeneous clinical records into a
consistent representation of the treatment site, diagnosis, plan information,
dose prescription and relevant structure context, reducing variation caused by
free-text formatting rather than clinical content. The structured prompt was
then encoded using a frozen BioMedBERT-large encoder~\cite{gu2021domain}, a
BERT-based language model pretrained on large-scale biomedical text, to obtain a
domain-specific semantic representation of the current registration sample. The
resulting 768-dimensional embedding $\mathbf{e}$ was passed through a two-layer
prompt adapter with GELU activation. The adapter converted the semantic
embedding into feature-wise linear modulation parameters, allowing clinical
information to condition the decoder without replacing the image-driven
registration backbone. For decoder levels $l\in\{1,2,3,4,5\}$, the modulation
parameters were defined as:
\begin{equation}
  (\gamma_l, \beta_l) =
  \text{Linear}_l\bigl(\text{PromptAdapter}(\mathbf{e})\bigr).
\end{equation}

Spatial context was derived from available source-side structure contours and
dose maps for the current treatment plan. The goal was to provide the decoder
with spatial priors that indicate not only where clinically important anatomy is
located, but also where registration errors are more likely to affect adaptive assessment. As
illustrated in Fig.~\ref{fig:method_framework}c, CTV contours were converted
into signed-distance maps and transformed into Gaussian proximity and
boundary-band features, preserving both the interior of the CTV and
its clinically sensitive boundary. OAR contours were projected into binary
structure maps to identify organs in which contour-transfer or dose-mapping
errors would require clinical review. These contour-derived features were fused
to form a continuous anatomical proximity map
$\AnatMap\in[0,1]^{H\times W\times D}$. Dose maps were normalised and
processed to extract isodose-band and local dose-gradient features, because
small spatial errors near steep dose gradients can lead to disproportionate
dosimetric errors. These dose-derived features were combined with the OAR map to
emphasise dose-sensitive anatomy, yielding a dose-sensitive map
$\DosePriorMap\in[0,1]^{H\times W\times D}$. The anatomical proximity map
and dose-sensitive map were then combined to form a risk map:
\begin{equation}
  \RiskMap=1-(1-\AnatMap)(1-\DosePriorMap).
\end{equation}
The spatial prior
$\SpatialPriorMap=\text{concat}(\AnatMap,\RiskMap)$ was downsampled to each
decoder scale $l\in\{1,2,3,4,5\}$ and used as a spatial gate:
\begin{equation}
  \tilde{\mathbf{F}}_l=\mathbf{F}_l\odot
  \sigma\bigl(W_l[\mathbf{F}_l,\text{Pool}_l(\SpatialPriorMap)]+b_l\bigr),
\end{equation}
where $\sigma$ denotes the sigmoid function and $\odot$ denotes element-wise
multiplication. Semantic modulation was then applied to the spatially gated
decoder feature map:
\begin{equation}
  \hat{\mathbf{F}}_l=\gamma_l\odot\tilde{\mathbf{F}}_l+\beta_l .
\end{equation}

\noindent\textbf{Patient-history conditioning.}
Patient-history conditioning was used only in Stage~2, while the current CT pair
and current multimodal clinical-context inputs from Stage~1 were retained. For a
current registration, the available patient history included previous
treatment-state images, previous Stage~1 deformation outputs and
registration-quality information from the same patient.
Together, these sources describe the anatomical states already observed, the
patient's prior motion and shape-change pattern, and the reliability of each
historical registration when updating the patient-specific memory.

To incorporate these sources into the current registration, Stage~2 converted
them into the memory representation shown in Fig.~\ref{fig:method_framework}d.
Directly processing the original historical CT volumes and full dense
deformation fields would substantially increase memory and computation and would
make variable-length patient histories difficult to handle. Therefore, each
previous time point was represented by a compact memory feature. After Stage~1
training, the trained Stage~1 model was applied to each previous source-target CT
pair from the same patient, and the resulting image features,
deformation-field statistics and quality measurements were stored as the history
representation.
For a patient with $h$ previous time points, the history sequence was indexed in
temporal order as $1,\ldots,h$. For each historical registration, the
Stage~1 encoder produced a 128-dimensional image feature vector $\mathbf{v}_t$
by global average pooling of bottleneck features. Deformation memory was
represented by four statistics, including mean displacement magnitude,
95th-percentile displacement, negative Jacobian fraction and Jacobian standard
deviation, forming
$\mathbf{d}_t\in\mathbb{R}^{4}$. These statistics summarised the magnitude and
regularity of the previous deformation without storing the full dense vector
field. Quality memory acted as a confidence summary and included
NCC$_\text{body}$, NCC$_\text{OAR}$, SSIM$_\text{body}$, SSIM$_\text{OAR}$ and
mean inverse-consistency error, forming $\mathbf{q}_t\in\mathbb{R}^{5}$.
Historical OAR-specific quality measurements were computed within source-derived
OAR masks warped into target space using the corresponding Stage~1 deformation
fields, without access to manual target contours. The per-time-point memory
feature was
\begin{equation}
  \mathbf{f}_t=[\mathbf{v}_t;\mathbf{d}_t;\mathbf{q}_t]\in\mathbb{R}^{137}.
\end{equation}
A trend feature was computed from successive deformation statistics:
\begin{equation}
  \boldsymbol{\tau}_t=\mathbf{d}_t-\mathbf{d}_{t-1},
\end{equation}
with $\boldsymbol{\tau}_1$ set to zero. The sequence
$(\mathbf{f}_1,\ldots,\mathbf{f}_h)$ was passed to a single-layer GRU with hidden
dimension 256. The GRU was used because it can summarise temporally ordered
histories with variable length while preserving the order of previously observed
anatomical changes, yielding the patient-history embedding
\begin{equation}
  \mathbf{m}=\text{GRU}(\mathbf{f}_1,\ldots,\mathbf{f}_h)_h .
\end{equation}
The history embedding was concatenated with the current source CT feature
$\mathbf{p}\in\mathbb{R}^{128}$ and current trend feature
$\boldsymbol{\tau}_h$:
\begin{equation}
  \mathbf{z}=[\mathbf{p};\mathbf{m};\boldsymbol{\tau}_h]\in\mathbb{R}^{388}.
\end{equation}
A two-layer memory adapter mapped $\mathbf{z}$ to patient-adaptive FiLM
parameters $(\gamma_l^*,\beta_l^*)$ for the same decoder levels
$l\in\{1,2,3,4,5\}$ used by the clinical-context module. These parameters provided
patient-specific conditioning in addition to the current clinical-context
conditioning.

\noindent\textbf{Two-stage training and objective.}
CAP-DIR was trained sequentially. Stage~1 was trained first using independent
current source-target CT pairs and the corresponding source-derived
clinical-context inputs, including structured text prompts, structure contours
and dose maps. In this stage, the shared registration backbone and
clinical-context conditioning
modules were optimised to estimate $\phi_{s\rightarrow t}$ for the current image
pair.

After Stage~1 convergence, historical representations were generated for the
training patients as described above. Stage~2 was then trained with
current-pair-plus-history sequences. For each current target image, the input
included the current CT pair, current clinical context and only previous time
points from the same patient. During Stage~2 training, the Stage~1 registration
backbone was initialised from the trained Stage~1 model. The image encoder was
kept frozen, whereas the decoder, GRU memory encoder, memory adapter and
patient-adaptive modulation parameters were optimised. During inference, patient
history was used only as input to the trained memory encoder, and no
patient-specific fine-tuning or test-time parameter update was performed.

Both stages used the same image-similarity and regularisation objective:
\begin{equation}
  \Loss =
  \Loss_\mathrm{NCC} +
  \lambda_\mathrm{OAR}\Loss_\mathrm{NCC}^{\mathrm{OAR}} +
  \lambda_\mathrm{reg}\Loss_\mathrm{reg}.
\end{equation}
$\Loss_\mathrm{NCC}$ and
$\Loss_\mathrm{NCC}^{\mathrm{OAR}}$ denote local NCC losses computed in the
foreground body and OAR regions, respectively. For
$\Loss_\mathrm{NCC}^{\mathrm{OAR}}$, the source-derived OAR mask was warped into
target space using the deformation field predicted for the current image pair.
The resulting mask provided a coarse target-space OAR localisation for computing
the OAR-weighted NCC loss, without using manual target contours. The foreground
body mask was used to prevent similarity optimisation from being dominated by padded background
voxels, while the OAR-weighted term increased the contribution of clinically
important structures. $\Loss_\mathrm{reg}$ is a diffusion regularisation
term on the spatial gradient of the displacement field.
Unless otherwise specified, $\lambda_\mathrm{OAR}=1.0$ and
$\lambda_\mathrm{reg}=0.2$.

\subsection{Implementation and comparison methods}

\noindent\textbf{Implementation details.}
All models were implemented in PyTorch 2.5.1 with CUDA 11.8. CAP-DIR training
was run with distributed data parallel optimisation on $3\times$ NVIDIA
A100-80G GPUs. Stage~1 used a batch size of 1 per GPU, 2,000 optimisation steps
per epoch. Optimisation was performed with Adam using a fixed
learning rate of $10^{-4}$ and $\lambda_\text{reg}=0.2$. The clinical-text
embedding dimension was 768, the prompt-modulation scale was initialised to 0.1,
and the spatial-prior weight was initialised to 0.1. These two scaling factors
allowed text-based modulation and contour-dose spatial priors to enter the
decoder gradually at the beginning of training, rather than dominating the
image-matching objective from the first optimisation steps. Stage~2 was
initialised from the Stage~1 checkpoint and optimised with the same learning
rate, batch size and regularisation weight. The Stage~1 and Stage~2 checkpoints
used for evaluation were selected independently according to their best
performance on the internal validation set. Historical information was
encoded with a GRU memory aggregator, and samples with longer histories were
up-weighted using a curriculum factor of $1+0.1h$, where $h$ is the number of
available previous time points. This weighting encouraged the memory encoder to
learn from patients with richer longitudinal histories while still retaining
short-history samples during training. All CAP-DIR variants and learning-based
comparison methods used the same preprocessed image pairs and patient-level data
splits.

\noindent\textbf{Comparison methods.}
CAP-DIR was compared with representative learning-based and conventional DIR
methods. VoxelMorph~\cite{balakrishnan2019voxelmorph} is a convolutional
unsupervised registration network. TransMorph~\cite{chen2022transmorph} uses a
transformer architecture to model long-range correspondence.
CorrMLP~\cite{wang2024corrmlp} introduces correlation-aware MLP blocks for
dense displacement estimation. MambaMorph~\cite{wang2024mambamorph} uses
state-space sequence modelling for medical image registration.
DiffuseMorph~\cite{kim2022diffusemorph} formulates registration through a
diffusion-based deformation generation process.
UniGradICON~\cite{tian2024unigradicon} is a general-purpose pretrained
registration model based on inverse-consistency regularisation. UniGradICON was
used directly without retraining, whereas the other learning-based baselines
were trained using the same preprocessed image pairs and patient-level data
partition as CAP-DIR. Direct transfer without registration was included to
quantify the uncorrected anatomical discrepancy. Demons registration was also
included because it is a conventional intensity-based DIR algorithm commonly
used in clinical image-processing systems, providing a reference for routine
clinical registration performance.

CAP-DIR variants were used for ablation analyses. The image-only variant used
the CT image pair without clinical-context conditioning. The clinical-context
variant added semantic and spatial clinical information. The full model added
patient-history adaptation to the clinical-context model and is denoted Ours w/
Clin.+Hist. in the figures. All Stage~1 clinical-context ablations used the same
architecture, training schedule, data partition and OAR-weighted objective,
differing only in the clinical-context inputs provided to the model.

\subsection{Blinded clinical contour review}

A retrospective method-blinded clinical review was conducted using 20
longitudinal planning-CT cases, with four cases each from the abdomen, brain,
breast/chest wall, head and neck, and pelvis. Each case included the
clinician-approved target contour set and two unedited contour sets propagated
by CAP-DIR and CorrMLP. The clinician-approved set was identified as the
reference, whereas the identities of the two propagation methods were concealed
using case-specific anonymous labels. All CT images and contour sets were
imported into the clinical ARIA environment at Mayo Clinic Arizona. Two
independent radiation oncology reviewers inspected the contours using the image
display and contour-review interface used in routine radiotherapy practice,
thereby approximating the clinical environment in which propagated contours
would be reviewed and edited. Reviewers rated CTVs and OARs separately on a
four-point ordinal scale: 4, clinically
acceptable without modification; 3, requiring minor edits; 2, requiring major
edits; and 1, unusable. The reviewers also estimated the time required to bring
the CTV and OAR contours to clinical usability. All three contour sets were
resampled to a common $2.5\times2.5\times4$-mm grid before review, and neither
automatically propagated set underwent manual correction before assessment.

\subsection{Evaluation metrics}

Registration performance was evaluated with image-similarity,
deformation-field and displacement metrics. Image similarity was measured with
local normalised cross-correlation (NCC) using a $9\times9\times9$ window and
structural similarity index (SSIM). Deformation-field plausibility was assessed
using Jacobian determinant statistics, including Jacobian deviation, and folding
ratio, defined as the fraction of voxels with a negative Jacobian determinant.
Deformation magnitude was measured as the mean displacement-field magnitude.

Contour propagation was evaluated with Dice similarity coefficient (DSC) and
residual symmetric-difference volume. DSC measured the overlap between
propagated contours and manual target contours. Residual volume was calculated
from the symmetric difference between each propagated contour and the manual
target contour and reported in millilitres; it is a geometric error measure and
not a direct measurement of clinician editing effort.

For the blinded clinical review, clinical acceptability was defined for analysis
as a rating of 3 or 4, indicating that a contour required no more
than minor edits. The clinician-based endpoints were the proportion of contours
meeting this threshold and the estimated editing time for CTVs and OARs. The
absolute difference in clinical acceptability and the paired reduction in
estimated editing time were used to quantify the benefit of CAP-DIR relative to
CorrMLP.

Dose accumulation was evaluated with dose mean absolute error (dose MAE) and
dose-volume histogram (DVH) endpoint error. Dose MAE measured the voxel-wise
difference between estimated and reference accumulated dose. DVH endpoint error
measured differences in clinically relevant dose-summary endpoints, including
Dmean and Dmax.

\subsection{Statistical analysis}

All quantitative analyses were performed on matched evaluation units, so that
each method was evaluated on the same registration case, contour-propagation
case or dose-accumulation case. Continuous metrics are reported as mean values with
standard deviations unless otherwise specified. For metrics where higher values
indicate better performance, relative improvement was calculated as the
difference from the comparator divided by the comparator value. For error metrics
where lower values indicate better performance, relative reduction was calculated
as the comparator value minus the CAP-DIR value, divided by the comparator
value. Paired comparisons between methods used two-sided Wilcoxon signed-rank
tests when statistical significance was assessed. Anatomical-site and external
cohort analyses were computed within each stratum using the matched cases
available for that stratum.

Clinical-review outcomes were analysed as paired observations matched by case
and reviewer. Ordinal ratings are reported as counts and proportions, with the
primary presentation based on the clinical-acceptability threshold.
The two reviewer estimates of editing time were averaged within each case before
reporting the median and interquartile range across cases. Absolute and relative
reductions were calculated from these case-level summaries. Confidence intervals
for differences in clinical acceptability were obtained by patient-level
bootstrap resampling. Inter-reviewer agreement for ordinal ratings was assessed
using linearly weighted Cohen's $\kappa$.


\section*{Data Availability}
The institutional radiotherapy cohorts used in this study are not publicly
available because they contain protected clinical imaging and treatment
information covered by patient privacy regulations, institutional review board
approval and data-use agreements. Subject to future institutional approval, a
small number of fully de-identified example cases may be released or shared
through a controlled request process to support reproducibility. The public
benchmark datasets used for external validation are available from their
original release pages, including
\href{https://fdat.uni-tuebingen.de/records/75kj1-64747}{LongCT-Lung},
\href{https://www.cancerimagingarchive.net/analysis-result/nlst-new-lesion-longct/}{NLST-New-lesion-LongCT},
\href{https://www.cancerimagingarchive.net/collection/pancreatic-ct-cbct-seg/}{Pancreas-CT},
\href{https://www.cancerimagingarchive.net/collection/4d-lung/}{4D-Lung}
and
\href{https://www.cancerimagingarchive.net/collection/rider-lung-ct/}{RIDER Lung CT}.
Users should comply with the access conditions and data-use policies of each
source dataset.

\section*{Code Availability}
The code for data preprocessing, model training, testing and evaluation will be
released on GitHub upon publication. The repository will include the
preprocessing scripts, CAP-DIR training code, baseline training and evaluation
pipelines, configuration files and instructions for reproducing the main
experiments. Trained model parameters derived from the institutional datasets
will not be publicly released because of institutional data-governance
requirements.

\section*{Acknowledgements}
This work was supported in part by the Department of Radiation Oncology, Mayo Clinic, Rochester, Minnesota, and Phoenix, Arizona, USA, R01 CA 261932 and R01 CA 272602 to Robert W. Mutter, and P30 CA015083 to MCCCC. This research was also supported by NIH/NIBIB R01EB293388, by NIH/NCI R01CA280134, by the Eric \& Wendy Schmidt Fund for AI Research \& Innovation, and by the Kemper Marley
Foundation to Wei Liu.

\section*{Author contributions statement}
C.J. and W.L. conceived the study. C.J., S.Z., Y.D. and M.J. contributed to
method development, software implementation, data preparation, experiments and
analysis. A.B. and K.K.L. performed the blinded clinical contour review. S.H.P.,
T.T.S., J.B.A., L.A.M., J.-C.M.R., W.G.R., S.R.K., S.A.V., W.W.W., N.Y.Y.,
M.Y.H. and S.E.S. provided clinical expertise and contributed to study design
and interpretation of the results. D.S. and W.L. supervised the study, and W.L.
provided project administration and funding support. C.J. prepared the original
manuscript. All authors reviewed and edited the manuscript and approved the
final version.

\section*{Competing interests}
The authors declare no competing interests.


\bibliography{dir_paper}

\begin{thebibliography}{10}
\urlstyle{rm}
\expandafter\ifx\csname url\endcsname\relax
  \def\url#1{\texttt{#1}}\fi
\expandafter\ifx\csname urlprefix\endcsname\relax\def\urlprefix{URL }\fi
\expandafter\ifx\csname doiprefix\endcsname\relax\def\doiprefix{DOI: }\fi
\providecommand{\bibinfo}[2]{#2}
\providecommand{\eprint}[2][]{\url{#2}}

\bibitem{sonke2019adaptive}
\bibinfo{author}{Sonke, J.-J.}, \bibinfo{author}{Aznar, M.} \&
  \bibinfo{author}{Rasch, C.}
\newblock \bibinfo{journal}{\bibinfo{title}{Adaptive radiotherapy for
  anatomical changes}}.
\newblock {\emph{\JournalTitle{Seminars in Radiation Oncology}}}
  \textbf{\bibinfo{volume}{29}}, \bibinfo{pages}{245--257},
  \doiprefix\url{10.1016/j.semradonc.2019.02.006} (\bibinfo{year}{2019}).

\bibitem{glidehurst2021adaptive}
\bibinfo{author}{Glide-Hurst, C.~K.} \emph{et~al.}
\newblock \bibinfo{journal}{\bibinfo{title}{Adaptive radiation therapy ({ART})
  strategies and technical considerations: A state of the {ART} review from
  {NRG} oncology}}.
\newblock {\emph{\JournalTitle{International Journal of Radiation Oncology
  Biology Physics}}} \textbf{\bibinfo{volume}{109}},
  \bibinfo{pages}{1054--1075}, \doiprefix\url{10.1016/j.ijrobp.2020.10.021}
  (\bibinfo{year}{2021}).

\bibitem{hall2022magnetic}
\bibinfo{author}{Hall, W.~A.} \emph{et~al.}
\newblock \bibinfo{journal}{\bibinfo{title}{Magnetic resonance linear
  accelerator technology and adaptive radiation therapy: An overview for
  clinicians}}.
\newblock {\emph{\JournalTitle{CA: A Cancer Journal for Clinicians}}}
  \textbf{\bibinfo{volume}{72}}, \bibinfo{pages}{34--56},
  \doiprefix\url{10.3322/caac.21707} (\bibinfo{year}{2022}).

\bibitem{yan1997adaptive}
\bibinfo{author}{Yan, D.}, \bibinfo{author}{Vicini, F.}, \bibinfo{author}{Wong,
  J.} \& \bibinfo{author}{Martinez, A.}
\newblock \bibinfo{journal}{\bibinfo{title}{Adaptive radiation therapy}}.
\newblock {\emph{\JournalTitle{Physics in Medicine and Biology}}}
  \textbf{\bibinfo{volume}{42}}, \bibinfo{pages}{123--132},
  \doiprefix\url{10.1088/0031-9155/42/1/008} (\bibinfo{year}{1997}).

\bibitem{lavrova2023adaptive}
\bibinfo{author}{Lavrova, E.} \emph{et~al.}
\newblock \bibinfo{journal}{\bibinfo{title}{Adaptive radiation therapy: A
  review of {CT}-based techniques}}.
\newblock {\emph{\JournalTitle{Radiology: Imaging Cancer}}}
  \textbf{\bibinfo{volume}{5}}, \bibinfo{pages}{e230011},
  \doiprefix\url{10.1148/rycan.230011} (\bibinfo{year}{2023}).

\bibitem{balakrishnan2019voxelmorph}
\bibinfo{author}{Balakrishnan, G.}, \bibinfo{author}{Zhao, A.},
  \bibinfo{author}{Sabuncu, M.~R.}, \bibinfo{author}{Guttag, J.} \&
  \bibinfo{author}{Dalca, A.~V.}
\newblock \bibinfo{journal}{\bibinfo{title}{{VoxelMorph}: {A} learning
  framework for deformable medical image registration}}.
\newblock {\emph{\JournalTitle{{IEEE} Transactions on Medical Imaging}}}
  \textbf{\bibinfo{volume}{38}}, \bibinfo{pages}{1788--1800},
  \doiprefix\url{10.1109/TMI.2019.2897538} (\bibinfo{year}{2019}).

\bibitem{chen2022transmorph}
\bibinfo{author}{Chen, J.} \emph{et~al.}
\newblock \bibinfo{journal}{\bibinfo{title}{{TransMorph}: Transformer for
  unsupervised medical image registration}}.
\newblock {\emph{\JournalTitle{Medical Image Analysis}}}
  \textbf{\bibinfo{volume}{82}}, \bibinfo{pages}{102615},
  \doiprefix\url{10.1016/j.media.2022.102615} (\bibinfo{year}{2022}).

\bibitem{wang2024corrmlp}
\bibinfo{author}{Meng, M.}, \bibinfo{author}{Feng, D.}, \bibinfo{author}{Bi,
  L.} \& \bibinfo{author}{Kim, J.}
\newblock \bibinfo{title}{Correlation-aware coarse-to-fine {MLP}s for
  deformable medical image registration}.
\newblock In \emph{\bibinfo{booktitle}{Proceedings of the {IEEE}/{CVF}
  Conference on Computer Vision and Pattern Recognition ({CVPR})}},
  \bibinfo{pages}{9645--9654}, \doiprefix\url{10.1109/CVPR52733.2024.00921}
  (\bibinfo{year}{2024}).

\bibitem{wang2024mambamorph}
\bibinfo{author}{Guo, T.} \emph{et~al.}
\newblock \bibinfo{journal}{\bibinfo{title}{{MambaMorph}: a {Mamba}-based
  framework for medical {MR}-{CT} deformable registration}}.
\newblock {\emph{\JournalTitle{arXiv preprint arXiv:2401.13934}}}
  \doiprefix\url{10.48550/arXiv.2401.13934} (\bibinfo{year}{2024}).

\bibitem{kim2022diffusemorph}
\bibinfo{author}{Kim, B.}, \bibinfo{author}{Han, I.} \& \bibinfo{author}{Ye,
  J.~C.}
\newblock \bibinfo{title}{{DiffuseMorph}: Unsupervised deformable image
  registration using diffusion model}.
\newblock In \emph{\bibinfo{booktitle}{Proceedings of the European Conference
  on Computer Vision ({ECCV})}}, vol. \bibinfo{volume}{13662} of
  \emph{\bibinfo{series}{Lecture Notes in Computer Science}},
  \bibinfo{pages}{347--364}, \doiprefix\url{10.1007/978-3-031-20086-1_20}
  (\bibinfo{publisher}{Springer, Cham}, \bibinfo{year}{2022}).

\bibitem{tian2024unigradicon}
\bibinfo{author}{Tian, L.} \emph{et~al.}
\newblock \bibinfo{title}{{uniGradICON}: A foundation model for medical image
  registration}.
\newblock In \emph{\bibinfo{booktitle}{Medical Image Computing and Computer
  Assisted Intervention -- {MICCAI} 2024}}, vol. \bibinfo{volume}{15002} of
  \emph{\bibinfo{series}{Lecture Notes in Computer Science}},
  \bibinfo{pages}{749--760}, \doiprefix\url{10.1007/978-3-031-72069-7_70}
  (\bibinfo{publisher}{Springer, Cham}, \bibinfo{year}{2024}).

\bibitem{perez2018film}
\bibinfo{author}{Perez, E.}, \bibinfo{author}{Strub, F.},
  \bibinfo{author}{de~Vries, H.}, \bibinfo{author}{Dumoulin, V.} \&
  \bibinfo{author}{Courville, A.}
\newblock \bibinfo{title}{{FiLM}: Visual reasoning with a general conditioning
  layer}.
\newblock In \emph{\bibinfo{booktitle}{Proceedings of the AAAI Conference on
  Artificial Intelligence}}, vol.~\bibinfo{volume}{32},
  \doiprefix\url{10.1609/aaai.v32i1.11671} (\bibinfo{year}{2018}).

\bibitem{singhal2023large}
\bibinfo{author}{Singhal, K.} \emph{et~al.}
\newblock \bibinfo{journal}{\bibinfo{title}{Large language models encode
  clinical knowledge}}.
\newblock {\emph{\JournalTitle{Nature}}} \textbf{\bibinfo{volume}{620}},
  \bibinfo{pages}{172--180}, \doiprefix\url{10.1038/s41586-023-06291-2}
  (\bibinfo{year}{2023}).

\bibitem{moor2023foundation}
\bibinfo{author}{Moor, M.} \emph{et~al.}
\newblock \bibinfo{journal}{\bibinfo{title}{Foundation models for generalist
  medical artificial intelligence}}.
\newblock {\emph{\JournalTitle{Nature}}} \textbf{\bibinfo{volume}{616}},
  \bibinfo{pages}{259--265}, \doiprefix\url{10.1038/s41586-023-05881-4}
  (\bibinfo{year}{2023}).

\bibitem{cho2014learning}
\bibinfo{author}{Cho, K.} \emph{et~al.}
\newblock \bibinfo{title}{Learning phrase representations using {RNN}
  encoder-decoder for statistical machine translation}.
\newblock In \emph{\bibinfo{booktitle}{Proceedings of the 2014 Conference on
  Empirical Methods in Natural Language Processing ({EMNLP})}},
  \bibinfo{pages}{1724--1734}, \doiprefix\url{10.3115/v1/D14-1179}
  (\bibinfo{publisher}{Association for Computational Linguistics},
  \bibinfo{year}{2014}).

\bibitem{thirion1998image}
\bibinfo{author}{Thirion, J.-P.}
\newblock \bibinfo{journal}{\bibinfo{title}{Image matching as a diffusion
  process: an analogy with {Maxwell}'s demons}}.
\newblock {\emph{\JournalTitle{Medical Image Analysis}}}
  \textbf{\bibinfo{volume}{2}}, \bibinfo{pages}{243--260},
  \doiprefix\url{10.1016/S1361-8415(98)80022-4} (\bibinfo{year}{1998}).

\bibitem{wasserthal2023totalsegmentator}
\bibinfo{author}{Wasserthal, J.} \emph{et~al.}
\newblock \bibinfo{journal}{\bibinfo{title}{{TotalSegmentator}: Robust
  segmentation of 104 anatomic structures in {CT} images}}.
\newblock {\emph{\JournalTitle{Radiology: Artificial Intelligence}}}
  \textbf{\bibinfo{volume}{5}}, \bibinfo{pages}{e230024},
  \doiprefix\url{10.1148/ryai.230024} (\bibinfo{year}{2023}).

\bibitem{appenzoller2012predicting}
\bibinfo{author}{Appenzoller, L.~M.}, \bibinfo{author}{Michalski, J.~M.},
  \bibinfo{author}{Thorstad, W.~L.}, \bibinfo{author}{Mutic, S.} \&
  \bibinfo{author}{Moore, K.~L.}
\newblock \bibinfo{journal}{\bibinfo{title}{Predicting dose--volume histograms
  for organs-at-risk in {IMRT} planning}}.
\newblock {\emph{\JournalTitle{Medical Physics}}}
  \textbf{\bibinfo{volume}{39}}, \bibinfo{pages}{7446--7461},
  \doiprefix\url{10.1118/1.4761864} (\bibinfo{year}{2012}).

\bibitem{gu2021domain}
\bibinfo{author}{Gu, Y.} \emph{et~al.}
\newblock \bibinfo{journal}{\bibinfo{title}{Domain-specific language model
  pretraining for biomedical natural language processing}}.
\newblock {\emph{\JournalTitle{{ACM} Transactions on Computing for
  Healthcare}}} \textbf{\bibinfo{volume}{3}}, \bibinfo{pages}{1--23},
  \doiprefix\url{10.1145/3458754} (\bibinfo{year}{2021}).

\end{thebibliography}
\onecolumn
\captionsetup{font=scriptsize,skip=3pt}
\section*{Figures}
\begin{center}
\centering
\safeincludegraphics[width=0.82\textwidth,height=0.84\textheight,keepaspectratio]{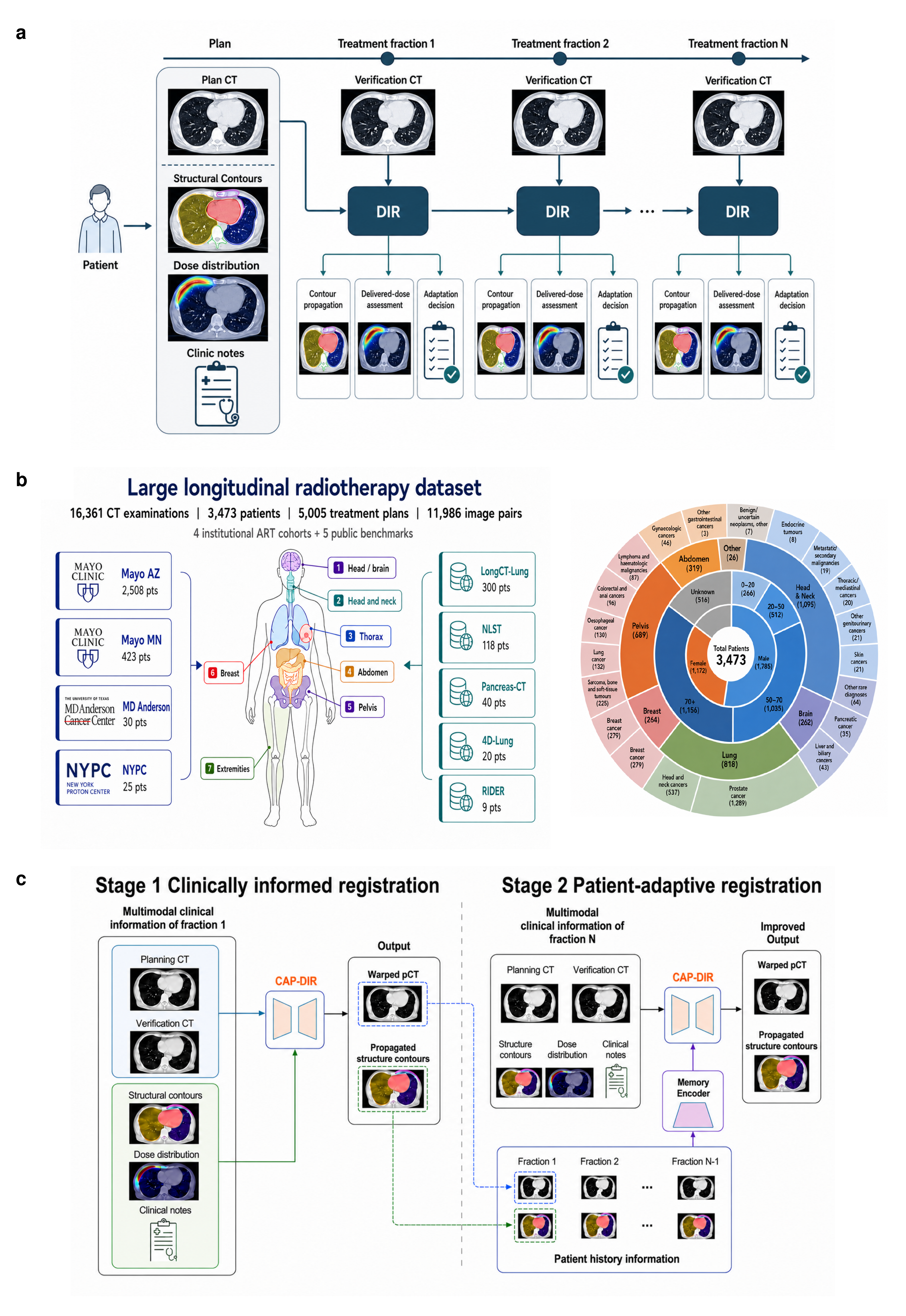}
\captionof{figure}{\textbf{Overview of clinical motivation, experimental data and method
design in this study.}
\textbf{a,} The longitudinal ART workflow from treatment planning to
fraction-level verification imaging, DIR-based contour propagation, delivered-dose
assessment and adaptation decisions.
\textbf{b,} The large-scale longitudinal radiotherapy dataset used for CAP-DIR
development and evaluation, including multi-institutional cohorts, public
benchmarks, anatomical coverage and patient composition.
\textbf{c,} The two-stage CAP-DIR workflow integrating clinically informed
registration with patient-adaptive memory from previous Stage~1 outputs.}
\label{fig:overview}
\end{center}
\clearpage

\begin{center}
\centering
  \safeincludegraphics[width=0.92\textwidth,height=0.84\textheight,keepaspectratio]{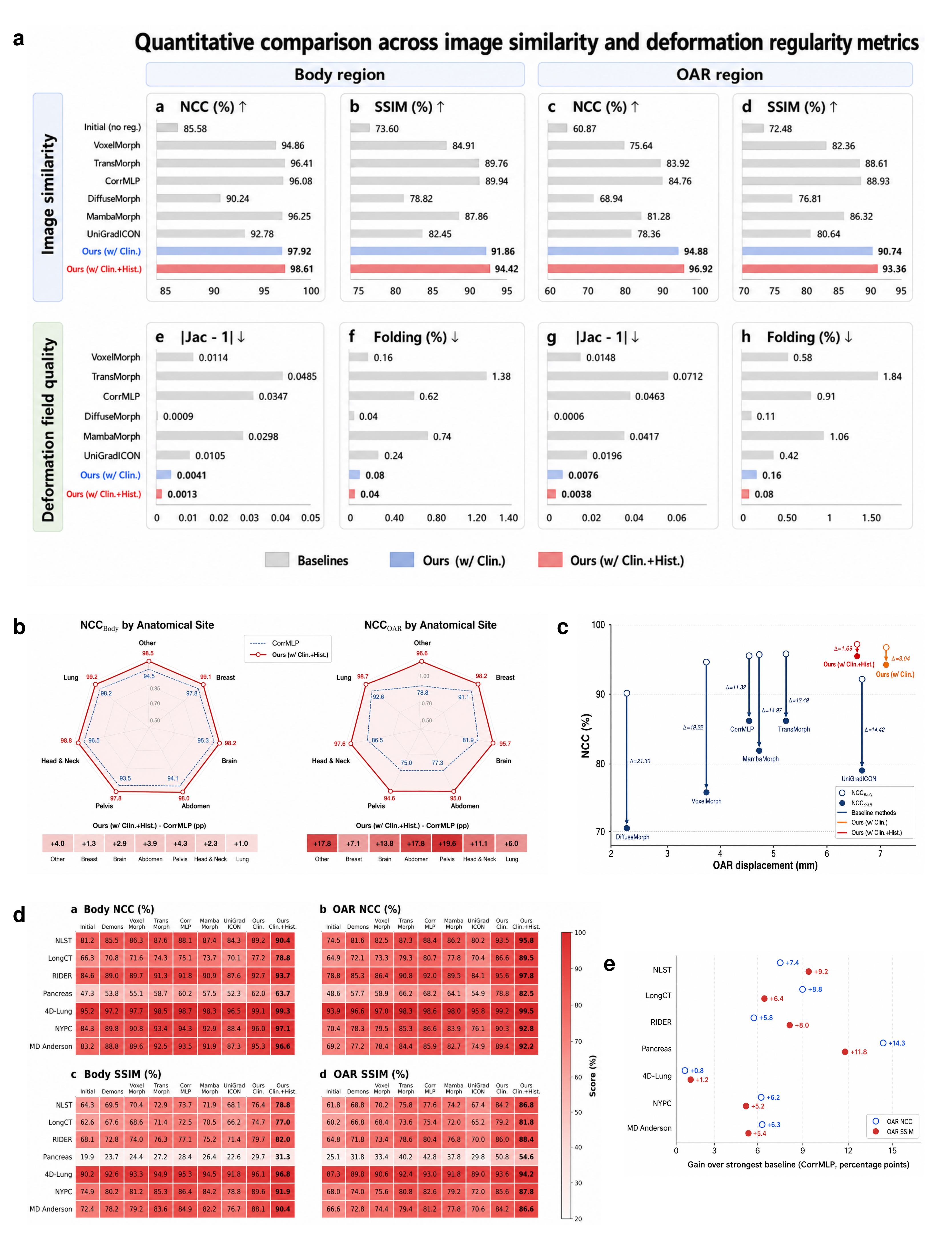}
  \captionof{figure}{\textbf{CAP-DIR improves OAR-focused registration performance across
    internal and external cohorts.}
    \textbf{a,} Quantitative comparison across image similarity and deformation
    regularity metrics on the internal test set.
    \textbf{b,} Body- and OAR-region gains over the strongest overall
    comparison method across anatomical sites.
    \textbf{c,} Relationship between Body-to-OAR performance drop and estimated
    OAR displacement magnitude.
    \textbf{d,} External validation heatmaps comparing all methods across held-out
    cohorts.
    \textbf{e,} OAR-region gains over the strongest comparison method across
    external validation cohorts.}
  \label{fig:registration_performance}
\end{center}
\clearpage

\begin{center}
\centering
  \safeincludegraphics[width=0.82\textwidth,height=0.40\textheight,keepaspectratio,
    trim=0 228.6bp 0 0,clip]{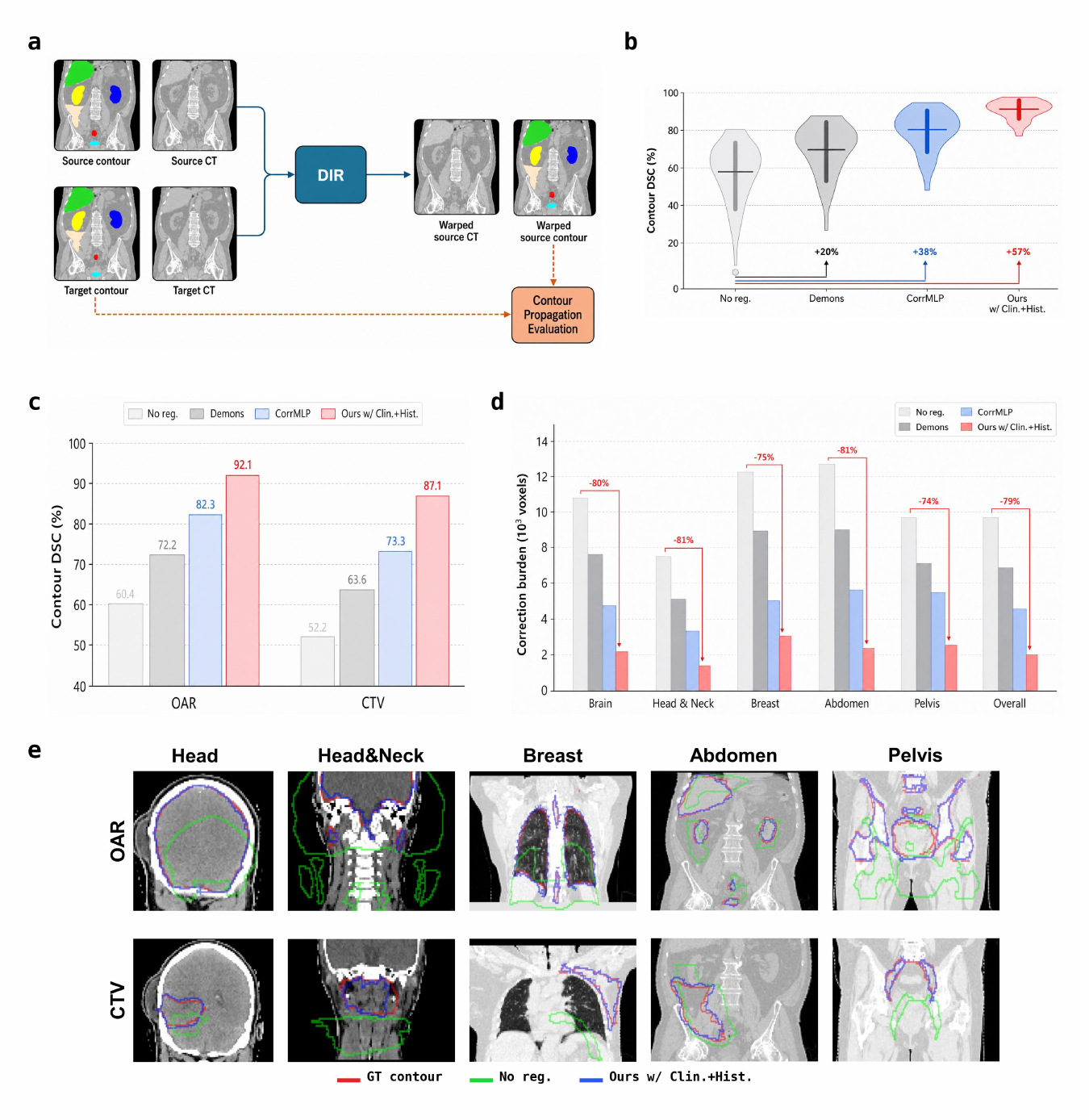}
  \par\vspace{0.25em}
  \safeincludegraphics[width=0.82\textwidth,height=0.22\textheight,keepaspectratio]{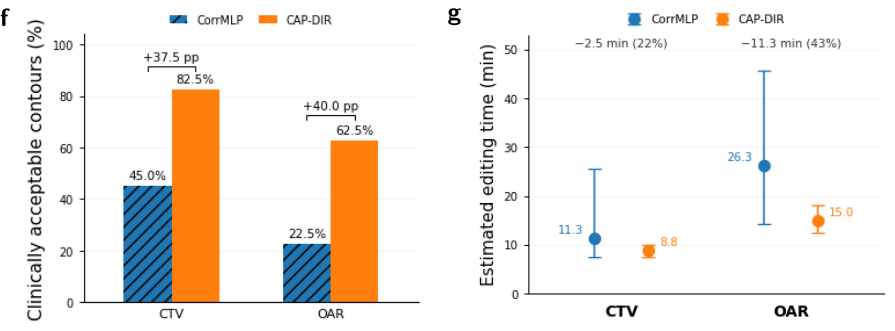}
  \par\vspace{0.25em}
  \safeincludegraphics[width=0.88\textwidth,height=0.22\textheight,keepaspectratio,
    trim=0 0 0 437.4bp,clip]{./figure/2.3/Figure3_contour_propagation_combined.pdf}
  \captionof{figure}{\textbf{Improved registration translates into more accurate contour
    propagation.}
    \textbf{a,} Evaluation design using longitudinal planning CT pairs with
    manual contours available on both source and target images.
    \textbf{b,} Overall contour DSC for direct transfer, Demons, CorrMLP and
    CAP-DIR.
    \textbf{c,} OAR- and CTV-specific contour DSC.
    \textbf{d,} Residual symmetric-difference volume across anatomical sites;
    $10^3$ voxels on the 1-mm$^3$ evaluation grid correspond to 1~ml.
    \textbf{f,} Proportion of CTV and OAR contours rated clinically acceptable
    (no more than minor edits) in the blinded clinical review; each method
    comprised 40 reader--case evaluations.
    \textbf{g,} Median clinician-estimated editing time, with interquartile
    ranges, after averaging the two reviewer estimates within each case.
    \textbf{e,} Representative OAR and CTV propagation examples across
    anatomical sites.}
  \label{fig:contour}
\end{center}
\clearpage

\begin{center}
\centering
  \safeincludegraphics[width=0.96\textwidth,height=0.79\textheight,keepaspectratio]{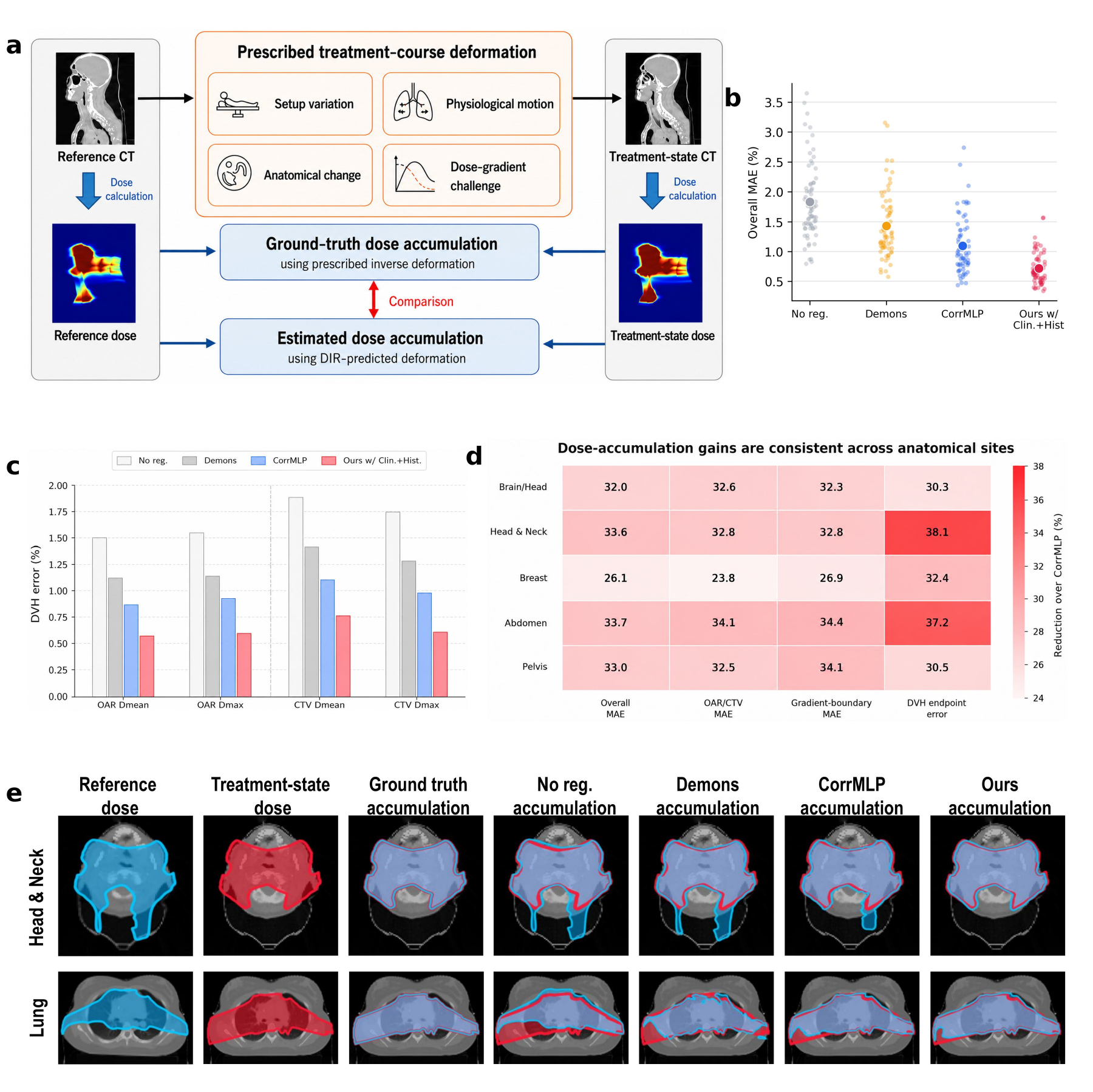}
  \captionof{figure}{\textbf{Improved registration supports dose accumulation evaluation.}
    \textbf{a,} Evaluation design for simulated dose accumulation. A prescribed
    treatment-course deformation generates a treatment-state CT and dose from a
    reference CT-dose pair. Reference accumulation adds the reference dose to the
    treatment-state dose mapped by the prescribed inverse deformation, whereas
    estimated accumulation adds the reference dose to the treatment-state dose
    mapped by direct transfer or DIR-predicted deformation.
    \textbf{b,} Overall accumulated-dose mean absolute error (MAE) for direct
    transfer without registration, Demons, CorrMLP and CAP-DIR.
    \textbf{c,} Absolute DVH endpoint error for OAR and CTV Dmean and Dmax.
    \textbf{d,} Anatomical-site consistency of CAP-DIR error reduction relative
    to CorrMLP across global, structure-region, gradient-boundary and DVH
    endpoint metrics.
    \textbf{e,} Representative head-and-neck and lung visualisations comparing
    reference dose, treatment-state dose, reference accumulation and accumulated
    dose estimated by each registration method.}
  \label{fig:dose}
\end{center}
\clearpage

\begin{center}
\centering
  \safeincludegraphics[width=0.92\textwidth,height=0.85\textheight,keepaspectratio]{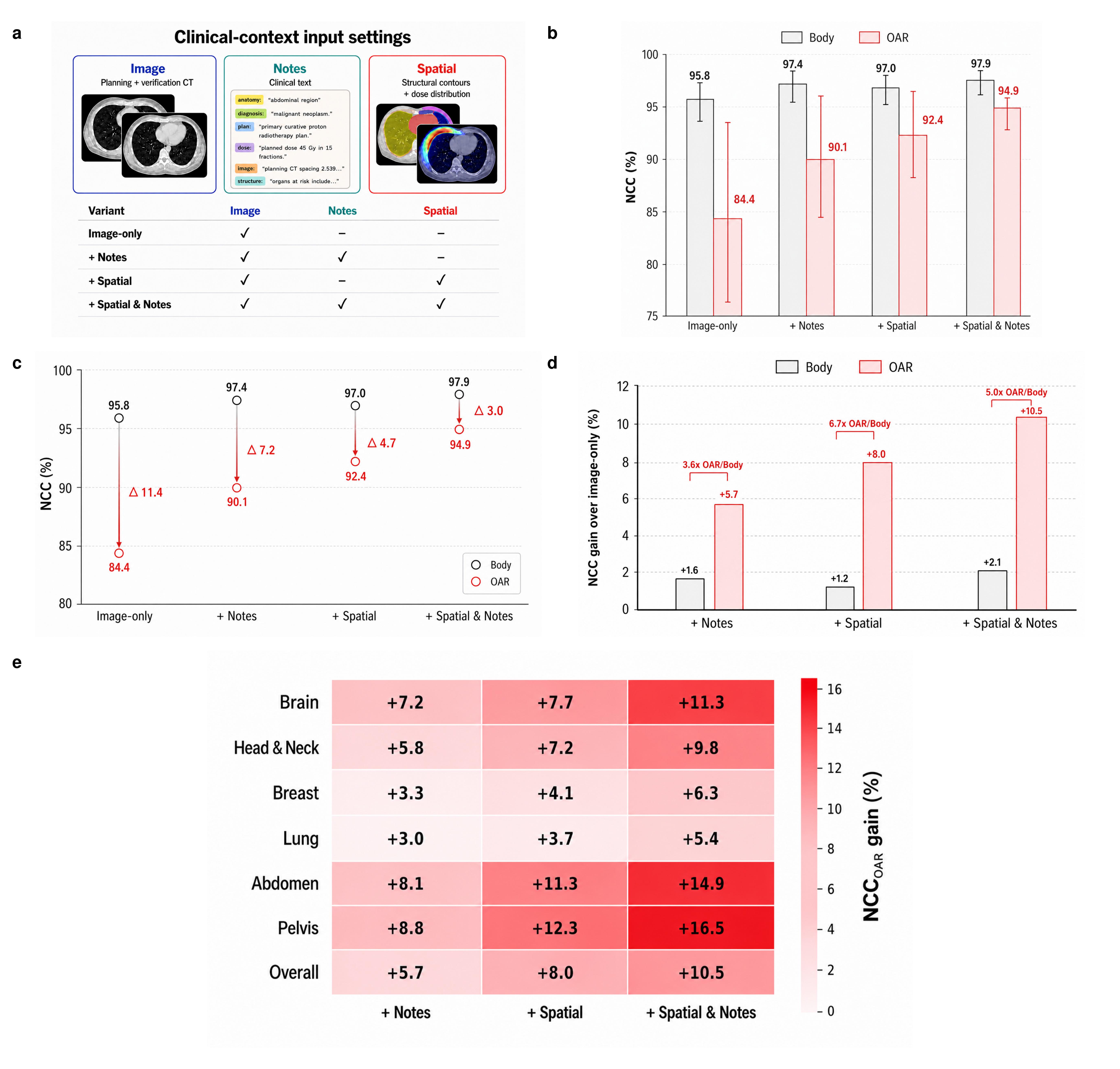}
  \captionof{figure}{\textbf{Multimodal clinical context improves registration and redirects
    gains to clinically relevant regions.}
    \textbf{a,} Ablation design comparing image-only registration with semantic
    clinical notes, spatial clinical context, and their combination.
    \textbf{b,} Body and OAR-region NCC across clinical-context variants.
    \textbf{c,} Body-to-OAR NCC gap across the same variants.
    \textbf{d,} Body and OAR-region NCC gains over the image-only model,
    highlighting preferential improvement in clinically important regions.
    \textbf{e,} Anatomical-site heatmap of OAR-region NCC gain relative to
    image-only registration.}
  \label{fig:clinical_context_effect}
\end{center}
\clearpage

\begin{center}
\centering
  \safeincludegraphics[width=0.95\textwidth,height=0.85\textheight,keepaspectratio]{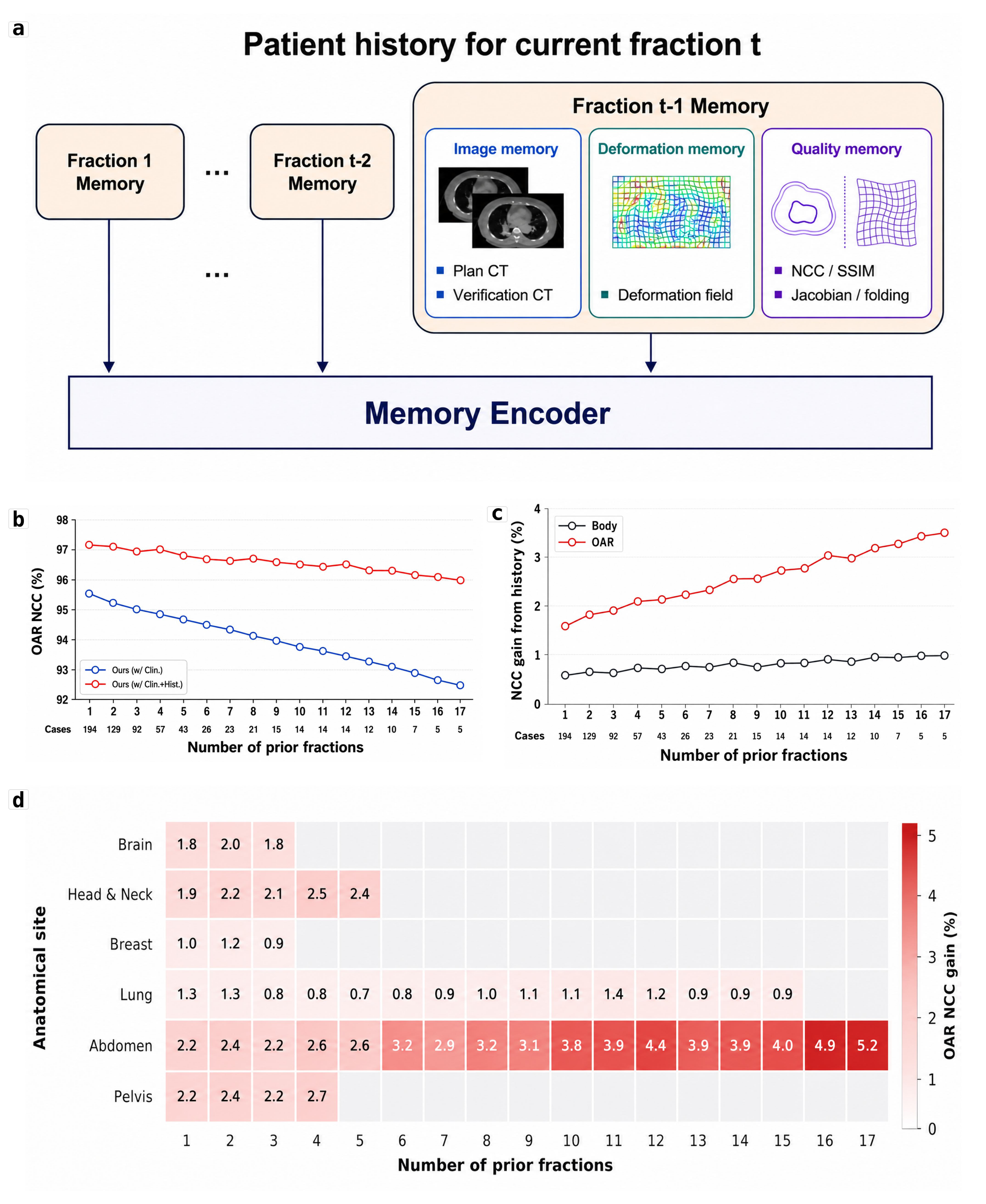}
  \captionof{figure}{\textbf{Patient-history adaptation improves registration as
    longitudinal evidence accumulates.}
    \textbf{a,} Patient history available for a current treatment fraction,
    including image, deformation-field, and registration-quality memories from
    previous fractions, which are summarised by a memory encoder.
    \textbf{b,} OAR-region NCC as a function of the number of prior fractions for
    CAP-DIR with clinical context alone and with clinical context plus patient
    history.
    \textbf{c,} NCC gain from patient history in the Body and OAR regions across
    increasing history lengths.
    \textbf{d,} Anatomical-site heatmap of OAR-region NCC gain from patient
    history across history lengths.}
  \label{fig:adaptive}
\end{center}
\clearpage

\begin{center}
\centering
  \safeincludegraphics[width=0.92\textwidth,height=0.77\textheight,keepaspectratio]{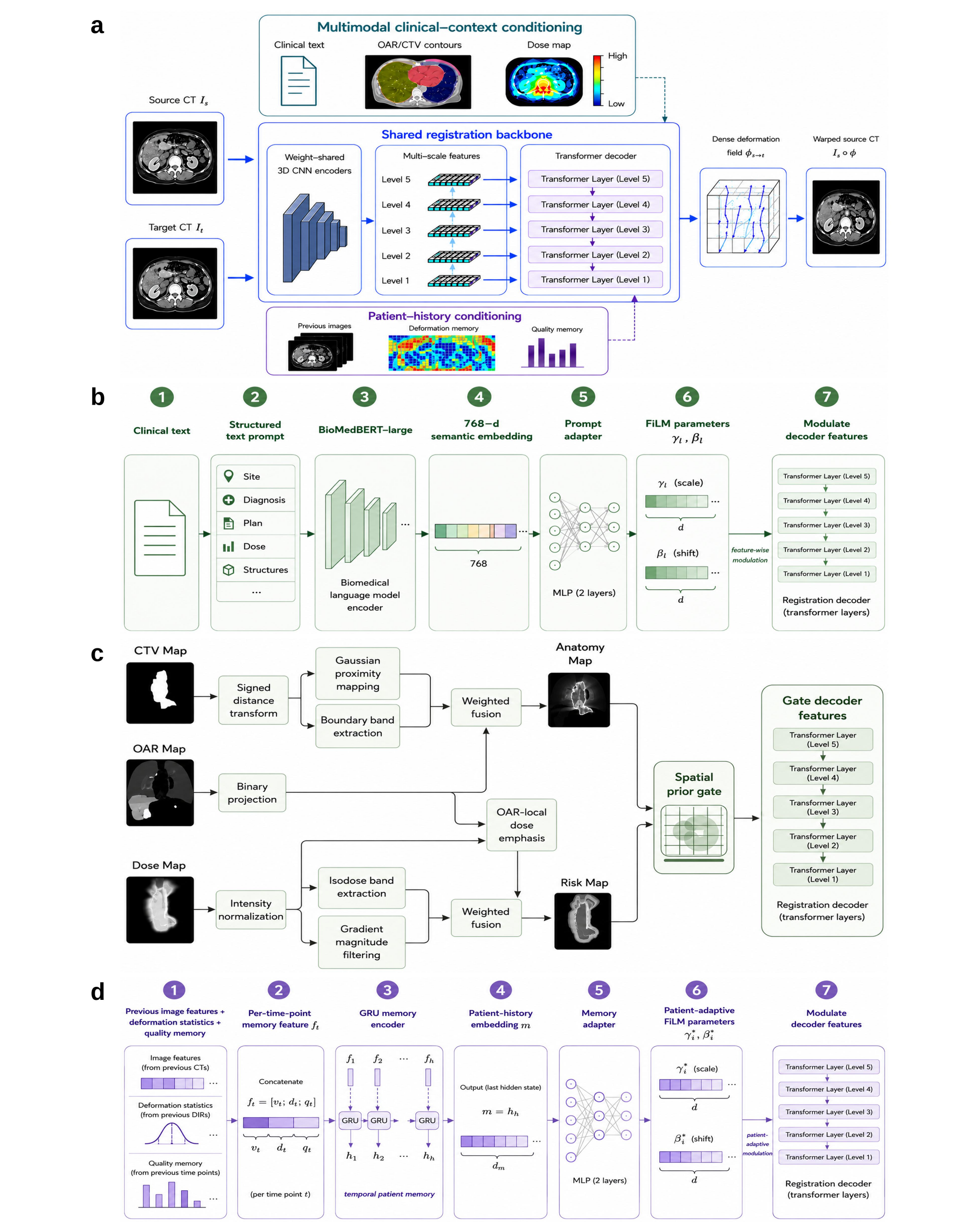}
  \captionof{figure}{\textbf{CAP-DIR conditional registration framework.}
    \textbf{a,} Overall CAP-DIR architecture. Source and target CTs are encoded
    by weight-shared 3D CNNs, fused as multi-scale features and decoded by a
    coarse-to-fine transformer to estimate a dense deformation field for source
    CT warping. Multimodal clinical context conditions Stage~1, and
    patient-history information provides additional conditioning in Stage~2.
    \textbf{b,} Semantic clinical-context conditioning converts structured
    clinical text into BioMedBERT embeddings and FiLM parameters that modulate
    decoder features.
    \textbf{c,} Spatial clinical-context conditioning converts OAR/CTV contours
    and dose maps into anatomy and risk maps that gate decoder features.
    \textbf{d,} Patient-history conditioning encodes previous image features,
    deformation statistics and quality memory with a GRU to generate
    patient-adaptive FiLM parameters.}
  \label{fig:method_framework}
\end{center}
\clearpage

\section*{Tables}
\begin{center}
\captionof{table}{\textbf{Dataset split overview.}
  Patient, treatment-plan, CT-examination and image-pair counts are reported for
  each cohort across the training, validation, internal test and external
  validation splits.}
\renewcommand\arraystretch{1.25}
\label{tab:dataset_split}
\centering
\resizebox{\textwidth}{!}{%
\begin{tabular}{lllrrrr}
\toprule
\rowcolor{tablehead}
\textbf{Split} & \textbf{Dataset} & \textbf{Institution / Source} &
  \textbf{Patients} & \textbf{Plans} & \textbf{Total CTs} & \textbf{Image pairs} \\
\midrule
\multirow{2}{*}{Training} & Mayo AZ & Mayo Clinic Arizona & 1,756 & 2,472 & 9,491 & 7,019 \\
 & Mayo MN & Mayo Clinic Minnesota & 296 & 547 & 547 & 410 \\
\rowcolor{tablesubtotal}
\textit{Subtotal} & & & \textit{2,052} & \textit{3,019} & \textit{10,038} & \textit{7,429} \\
\midrule
\multirow{2}{*}{Validation} & Mayo AZ & Mayo Clinic Arizona & 251 & 353 & 1,356 & 1,003 \\
 & Mayo MN & Mayo Clinic Minnesota & 42 & 78 & 78 & 59 \\
\rowcolor{tablesubtotal}
\textit{Subtotal} & & & \textit{293} & \textit{431} & \textit{1,434} & \textit{1,062} \\
\midrule
\multirow{2}{*}{Internal test} & Mayo AZ & Mayo Clinic Arizona & 501 & 706 & 2,711 & 2,005 \\
 & Mayo MN & Mayo Clinic Minnesota & 85 & 156 & 156 & 117 \\
\rowcolor{tablesubtotal}
\textit{Subtotal} & & & \textit{586} & \textit{862} & \textit{2,867} & \textit{2,122} \\
\midrule
\rowcolor{tablesubtotal}
\textbf{Internal total} & & & \textbf{2,931} & \textbf{4,312} & \textbf{14,339} & \textbf{10,613} \\
\midrule
\multirow{7}{*}{External validation} & MD Anderson & MD Anderson Cancer Center & 30 & 53 & 53 & 23 \\
 & NYPC & New York Proton Center & 25 & 46 & 46 & 21 \\
 & LongCT-Lung & Public & 300 & 331 & 684 & 353 \\
 & NLST & Public & 118 & 122 & 244 & 122 \\
 & Pancreas-CT & Public & 40 & 40 & 120 & 80 \\
 & 4D-Lung & Public & 20 & 80 & 800 & 720 \\
 & RIDER & Public & 9 & 21 & 75 & 54 \\
\rowcolor{tablesubtotal}
\textit{Subtotal} & & & \textit{542} & \textit{693} & \textit{2,022} & \textit{1,373} \\
\midrule
\rowcolor{tablehead}
\textbf{Grand total} & & & \textbf{3,473} & \textbf{5,005} & \textbf{16,361} & \textbf{11,986} \\
\bottomrule
\end{tabular}%
}
\end{center}
\clearpage

\clearpage


\section*{Supplementary material}

# Supplementary Information: Personalising Deformable Image Registration with Clinical Context and Deformation History for Adaptive Radiotherapy

**Caiwen Jiang[1], Songlin Zhao[1], Yuzhen Ding[1], Jia Mi[3], Samir H. Patel[1], Alexandra Bennett[1], Keldon K. Lin[1], Terence T. Sio[1], Jonathan B. Ashman[1], Lisa A. McGee[1], Jean-Claude M. Rwigema[1], William G. Rule[1], Sameer R. Keole[1], Sujay A. Vora[1], William W. Wong[1], Nathan Y. Yu[1], Michele Y. Halyard[1], Steven E. Schild[1], Safia K. Ahmed[1], Daniel J. Ma[2], Robert L. Foote[2], Michelle A. Neben Wittich[2], Thomas J. Whitaker[4], Haibo Lin[5], Dinggang Shen[3], and Wei Liu[1,*]**

[1]**Department of Radiation Oncology, Mayo Clinic, Phoenix, Arizona, USA**
[2]**Department of Radiation Oncology, Mayo Clinic, Rochester, Minnesota, USA**
[3]**School of Biomedical Engineering, ShanghaiTech University, Shanghai, China**
[4]**Department of Radiation Physics, The University of Texas MD Anderson Cancer Center, Houston, Texas, USA**
[5]**New York Proton Center, New York, New York, USA**
[*]**Correspondence to Wei Liu (liu.wei@mayo.edu).**

## Supplementary Notes

### Contour propagation and residual contour volume

The contour-propagation analysis used longitudinal planning CTs because routine verification CTs rarely contained complete manual reference contours. The first planning CT was used as the source and each eligible subsequent planning CT as a target. The internal cohort provided 276 source–target pairs; MD Anderson and the New York Proton Center provided 23 and 21 additional pairs, respectively. Direct transfer produced the greatest residual symmetric-difference volume, which CAP-DIR reduced from 9.66 to 2.06 ml. This corresponded to reductions of 79% relative to direct transfer and 55% relative to CorrMLP.

### Controlled dose-mapping results

The dose-mapping benchmark used prescribed synthetic treatment-course deformations to provide a known reference mapping. The reference accumulated dose combined the reference dose with the treatment-state dose mapped by the prescribed inverse deformation. Estimated accumulation used direct transfer or the deformation predicted by each DIR method. CAP-DIR reduced whole-volume dose mean absolute error by 34.5% relative to CorrMLP and reduced the mean error across OAR and CTV mean- and maximum-dose endpoints by 33.3%. Across anatomical sites, the reduction relative to CorrMLP was 27.7–35.9% for global error, 26.6–36.0% within structure regions, 28.9–36.3% at gradient boundaries and 30.3–38.1% for dose-volume-histogram endpoints.

### History-length stratification

Among cases stratified by the amount of available prior history, the OAR-region gain from Stage 2 increased from 1.6 percentage points with one prior observation to 3.5 percentage points with 17 prior observations; the body-region gain ranged from 0.6 to 1.0 percentage points. Abdominal cases showed the largest observed site-specific increase.

## Supplementary Tables

**Supplementary Table 1. Quantitative comparison of registration methods on the internal test set (586 patients, 2,122 registration cases from Mayo AZ and Mayo MN).** One registration case corresponds to one source–target CT pair; clinical context and available patient history are model inputs and do not alter the case count. All metrics are reported as mean ± s.d. Image similarity: NCC and SSIM (%) computed over whole-body (Body) and OAR-masked (OAR) regions. Deformation regularity: mean Jacobian determinant (closer to 1.0 indicates better local volume preservation) and folding ratio (percentage of voxels with a non-positive Jacobian determinant; lower is better), each computed over Body and OAR regions. Deformation magnitude: mean displacement field magnitude (mm) over Body and OAR regions, reported descriptively without ranking. Initial (no reg.): before-registration image similarity; deformation metrics not applicable (—). Bold = best per metric; underline = second best (displacement columns excluded from ranking). CAP-DIR (Clin.) uses clinical context, and CAP-DIR (Clin.+Hist.) additionally uses patient history. Both variants are evaluated on the full internal test set; for cases without prior fraction history, CAP-DIR (Clin.+Hist.) falls back to Stage 1 behaviour.

| Method | Image Similarity | | | | Deformation Regularity | | | | Deformation Magnitude | |
|---|---|---|---|---|---|---|---|---|---|---|
| | Body | | OAR | | Body | | OAR | | Body / OAR | |
| | NCC(%) $\uparrow$ | SSIM(%) $\uparrow$ | NCC(%) $\uparrow$ | SSIM(%) $\uparrow$ | Jac $\rightarrow$1 | Fold% $\downarrow$ | Jac $\rightarrow$1 | Fold% $\downarrow$ | Disp. (mm) | Disp. (mm) |
| Initial (no reg.) | 85.58±13.97 | 73.60±10.38 | 60.87±29.36 | 72.48±15.99 | — | — | — | — | — | — |
| VoxelMorph[1] | 84.91±8.42 | 84.91±8.78 | 75.46±25.84 | 82.36±13.52 | 0.9886±0.0112 | 0.16±0.22 | 1.0148±0.0485 | 0.58±0.34 | 4.03±4.18 | 3.74±5.15 |
| TransMorph[2] | 96.41±3.52 | 89.76±5.68 | 83.92±18.74 | 88.61±8.66 | 1.0485±0.0346 | 1.38±1.72 | 1.0712±0.0594 | 1.84±1.89 | 5.42±6.32 | 5.18±6.21 |
| CorrMLP[3] | 96.08±5.21 | 89.94±6.08 | 84.76±17.62 | 88.93±8.41 | 1.0347±0.0215 | 0.62±1.31 | 1.0463±0.0528 | 0.91±1.48 | 5.36±4.82 | 4.62±4.76 |
| DiffuseMorph[4] | 90.24±11.35 | 78.82±9.71 | 68.94±25.12 | 76.81±13.96 | **0.9991**±0.0086 | **0.04**±0.08 | **1.0006**±0.0121 | <u>0.11</u>±0.19 | 2.90±0.53 | 2.30±0.83 |
| MambaMorph[5] | 96.25±5.86 | 87.86±7.42 | 81.28±20.66 | 86.32±10.08 | 1.0298±0.0294 | 0.74±1.18 | 1.0417±0.0492 | 1.06±1.35 | 5.08±5.96 | 4.76±5.72 |
| UniGradICON[6] | 92.78±10.64 | 82.45±9.13 | 78.36±23.71 | 80.64±16.85 | 1.0105±0.0478 | 0.24±0.18 | 1.0196±0.0665 | 0.42±0.36 | 8.12±9.86 | 6.82±9.74 |
| **CAP-DIR (Clin.)** | <u>97.92</u>±2.18 | <u>91.86</u>±3.94 | <u>94.88</u>±4.26 | <u>90.74</u>±4.12 | 1.0041±0.0069 | 0.08±0.07 | 1.0076±0.0132 | 0.16±0.12 | 6.94±11.25 | 7.13±11.18 |
| **CAP-DIR (Clin.+Hist.)** | **98.61**±1.04 | **94.42**±2.26 | **96.92**±2.58 | **93.36**±2.72 | <u>1.0013</u>±0.0037 | **0.04**±0.04 | <u>1.0038</u>±0.0081 | **0.08**±0.07 | 6.59±11.03 | 6.54±10.10 |

Best and second-best results are shown in **bold** and <u>underline</u>, respectively (displacement columns are descriptive, not ranked).

**Supplementary Table 2. Per-site quantitative comparison of all registration methods on the internal test set.** Results are reported for all eight methods across seven anatomical sites (586 patients, 2,122 registration cases from Mayo AZ and Mayo MN). One registration case corresponds to one source–target CT pair; clinical context and available patient history do not alter the case count. Values are reported as mean ± standard deviation. All metrics follow the same definition as Supplementary Table 1: NCC and SSIM (%) for image similarity; mean Jacobian determinant (→1) and folding ratio (%↓) for deformation regularity, each computed at Body and OAR levels. CAP-DIR (Clin.) uses clinical context, and CAP-DIR (Clin.+Hist.) additionally uses patient history. Bold = best per metric and site; underline = second best.

| Site | Registration cases | Method | Image Similarity | | | | Deformation Regularity | | | |
|---|---|---|---|---|---|---|---|---|---|---|
| | | | Body | | OAR | | Body | | OAR | |
| | | | NCC(%) ↑ | SSIM(%) ↑ | NCC(%) ↑ | SSIM(%) ↑ | Jac →1 | Fold% ↓ | Jac →1 | Fold% ↓ |
| Other | 16 | VoxelMorph[1] | 92.29±8.71 | 82.23±9.08 | 66.95±26.36 | 77.81±13.79 | 0.99±0.01 | 0.22±0.23 | 1.02±0.05 | 0.67±0.36 |
| | | TransMorph[2] | 95.21±3.58 | 87.82±5.78 | 76.88±18.73 | 85.36±8.65 | 1.06±0.04 | 1.88±1.79 | 1.08±0.06 | 2.39±1.97 |
| | | CorrMLP[3] | 94.49±5.39 | 88.08±6.29 | 78.83±17.98 | 86.10±8.58 | 1.04±0.02 | 0.96±1.38 | 1.06±0.06 | 1.29±1.56 |
| | | DiffuseMorph | 85.65±11.15 | 74.89±9.54 | 57.76±24.01 | 70.60±13.34 | **1.00**±0.01 | 0.07±0.08 | **1.00**±0.01 | 0.18±0.19 |
| | | MambaMorph[5] | 94.35±6.01 | 85.46±7.61 | 73.92±20.87 | 82.73±10.18 | 1.04±0.03 | 1.06±1.24 | 1.05±0.05 | 1.43±1.41 |
| | | UniGradICON[6] | 87.52±9.77 | 77.94±8.39 | 65.46±20.77 | 71.47±14.76 | 1.03±0.05 | 0.32±0.18 | 1.04±0.07 | 0.57±0.35 |
| | | **CAP-DIR (Clin.)** | 97.75±2.16 | 91.55±3.88 | 94.41±3.67 | 90.34±3.80 | **1.00**±0.01 | 0.09±0.06 | 1.01±0.01 | 0.17±0.10 |
| | | **CAP-DIR (Clin.+Hist.)** | **98.51**±0.96 | **94.20**±2.09 | **96.64**±2.24 | **93.09**±2.50 | **1.00**±0.00 | **0.04**±0.04 | **1.00**±0.01 | **0.09**±0.06 |
| Breast | 161 | VoxelMorph[1] | 97.61±6.96 | 87.78±7.25 | 84.92±21.05 | 87.22±11.01 | 0.99±0.01 | 0.10±0.19 | **1.00**±0.04 | 0.49±0.29 |
| | | TransMorph[2] | 97.70±2.86 | 91.83±4.62 | 91.45±14.95 | 92.09±6.91 | 1.04±0.03 | 0.85±1.43 | 1.06±0.05 | 1.25±1.57 |
| | | CorrMLP[3] | 97.78±4.30 | 91.93±5.02 | 91.09±14.35 | 91.95±6.85 | 1.03±0.02 | 0.26±1.10 | 1.04±0.05 | 0.50±1.24 |
| | | DiffuseMorph | 95.15±8.90 | 83.02±7.62 | 80.89±19.17 | 83.45±10.65 | **1.00**±0.01 | **0.01**±0.06 | **1.00**±0.01 | **0.04**±0.15 |
| | | MambaMorph[5] | 98.28±4.80 | 90.43±6.08 | 89.14±16.66 | 90.16±8.13 | 1.02±0.03 | 0.39±0.99 | 1.03±0.04 | 0.66±1.13 |
| | | UniGradICON[6] | 98.40±7.81 | 87.27±6.70 | 92.14±16.58 | 90.43±11.79 | 0.99±0.04 | 0.16±0.14 | **1.00**±0.05 | 0.26±0.28 |
| | | **CAP-DIR (Clin.)** | 98.70±1.72 | 93.31±3.10 | 97.07±2.93 | 92.60±3.03 | **1.00**±0.01 | 0.04±0.05 | **1.00**±0.01 | 0.10±0.08 |
| | | **CAP-DIR (Clin.+Hist.)** | **99.07**±0.77 | **95.43**±1.67 | **98.22**±1.79 | **94.60**±1.99 | **1.00**±0.00 | 0.02±0.03 | **1.00**±0.01 | 0.05±0.05 |
| Brain | 160 | VoxelMorph[1] | 93.62±8.27 | 83.61±8.63 | 71.44±25.03 | 80.16±13.10 | 0.99±0.01 | 0.19±0.22 | 1.02±0.05 | 0.62±0.34 |
| | | TransMorph[2] | 95.83±3.40 | 88.82±5.49 | 80.52±17.78 | 87.04±8.22 | 1.05±0.04 | 1.62±1.70 | 1.08±0.06 | 2.11±1.87 |
| | | CorrMLP[3] | 95.31±5.12 | 89.04±5.97 | 81.90±17.07 | 87.56±8.15 | 1.04±0.02 | 0.78±1.31 | 1.05±0.05 | 1.10±1.48 |
| | | DiffuseMorph | 88.02±10.59 | 76.92±9.06 | 63.54±22.80 | 73.81±12.67 | **1.00**±0.01 | **0.05**±0.08 | **1.00**±0.01 | 0.14±0.18 |
| | | MambaMorph[5] | 95.33±5.71 | 86.70±7.23 | 77.73±19.82 | 84.59±9.67 | 1.03±0.03 | 0.90±1.17 | 1.05±0.05 | 1.24±1.34 |
| | | UniGradICON[6] | 90.24±9.28 | 80.27±7.97 | 72.13±19.72 | 76.21±14.01 | 1.02±0.05 | 0.28±0.17 | 1.03±0.06 | 0.49±0.33 |
| | | **CAP-DIR (Clin.)** | 97.17±2.05 | 90.47±3.69 | 92.78±3.48 | 88.96±3.61 | 1.01±0.01 | 0.12±0.06 | 1.01±0.01 | 0.22±0.10 |
| | | **CAP-DIR (Clin.+Hist.)** | **98.17**±0.92 | **93.45**±1.99 | **95.67**±2.13 | **92.17**±2.37 | **1.00**±0.00 | 0.06±0.04 | 1.01±0.01 | **0.11**±0.06 |
| Abdomen | 195 | VoxelMorph[1] | 91.62±8.93 | 81.53±9.31 | 64.70±27.03 | 76.64±14.14 | 0.99±0.01 | 0.23±0.24 | 1.03±0.05 | 0.69±0.37 |
| | | TransMorph[2] | 94.90±3.67 | 87.32±5.92 | 75.06±19.20 | 84.51±8.87 | 1.06±0.04 | 2.01±1.84 | 1.09±0.06 | 2.53±2.02 |
| | | CorrMLP[3] | 94.08±5.53 | 87.60±6.45 | 77.30±18.43 | 85.37±8.80 | 1.04±0.02 | 1.05±1.41 | 1.06±0.06 | 1.39±1.60 |
| | | DiffuseMorph | 84.46±11.43 | 73.88±9.78 | 54.87±24.61 | 68.99±13.68 | **1.00**±0.01 | 0.07±0.08 | **1.00**±0.01 | 0.19±0.20 |
| | | MambaMorph[5] | 93.86±6.17 | 84.84±7.81 | 72.02±21.39 | 81.80±10.44 | 1.04±0.03 | 1.15±1.27 | 1.05±0.05 | 1.53±1.45 |
| | | UniGradICON[6] | 86.16±10.02 | 76.77±8.60 | 62.13±21.29 | 69.10±15.13 | 1.03±0.05 | 0.34±0.18 | 1.05±0.07 | 0.61±0.36 |
| | | **CAP-DIR (Clin.)** | 96.80±2.21 | 89.79±3.98 | 91.74±3.76 | 88.08±3.89 | 1.01±0.01 | 0.13±0.06 | 1.01±0.01 | 0.25±0.11 |
| | | **CAP-DIR (Clin.+Hist.)** | **97.95**±0.99 | **92.98**±2.15 | **95.05**±2.30 | **91.58**±2.56 | **1.00**±0.00 | **0.06**±0.04 | 1.01±0.01 | **0.13**±0.06 |
| Pelvis | 421 | VoxelMorph[1] | 90.62±9.26 | 80.49±9.65 | 61.33±28.02 | 74.87±14.66 | 0.98±0.01 | 0.25±0.25 | 1.03±0.06 | 0.73±0.38 |
| | | TransMorph[2] | 94.43±3.81 | 86.56±6.14 | 72.32±19.91 | 83.25±9.20 | 1.06±0.04 | 2.20±1.90 | 1.09±0.07 | 2.74±2.09 |
| | | CorrMLP[3] | 93.46±5.73 | 86.88±6.69 | 75.00±19.11 | 84.27±9.12 | 1.04±0.02 | 1.18±1.47 | 1.06±0.06 | 1.54±1.66 |
| | | DiffuseMorph | 82.68±11.85 | 72.35±10.14 | 50.53±25.52 | 66.58±14.18 | **1.00**±0.01 | 0.09±0.09 | **1.01**±0.01 | 0.22±0.21 |
| | | MambaMorph[5] | 93.13±6.39 | 83.91±8.09 | 69.17±22.18 | 80.41±10.82 | 1.04±0.03 | 1.27±1.31 | 1.06±0.06 | 1.67±1.50 |
| | | UniGradICON[6] | 84.12±10.39 | 75.02±8.92 | 57.13±22.07 | 65.55±15.69 | 1.04±0.05 | 0.36±0.19 | 1.05±0.07 | 0.67±0.37 |
| | | **CAP-DIR (Clin.)** | 96.53±2.29 | 89.30±4.13 | 91.00±3.90 | 87.45±4.04 | 1.01±0.01 | 0.15±0.06 | 1.02±0.01 | 0.27±0.11 |
| | | **CAP-DIR (Clin.+Hist.)** | **97.79**±1.02 | **92.64**±2.22 | **94.61**±2.38 | **91.17**±2.65 | **1.00**±0.00 | **0.07**±0.04 | **1.01**±0.01 | **0.14**±0.07 |
| Head & Neck | 669 | VoxelMorph[1] | 95.61±7.61 | 85.70±7.94 | 78.18±23.04 | 83.69±12.06 | 0.99±0.01 | 0.14±0.20 | 1.01±0.05 | 0.55±0.31 |
| | | TransMorph[2] | 96.76±3.13 | 90.33±5.05 | 85.98±16.37 | 89.56±7.56 | 1.05±0.03 | 1.23±1.57 | 1.07±0.06 | 1.68±1.72 |
| | | CorrMLP[3] | 96.55±4.71 | 90.48±5.50 | 86.50±15.71 | 89.76±7.50 | 1.03±0.02 | 0.52±1.21 | 1.04±0.05 | 0.80±1.36 |
| | | DiffuseMorph | 91.58±9.75 | 79.97±8.34 | 72.21±20.99 | 78.63±11.66 | **1.00**±0.01 | **0.03**±0.07 | **1.00**±0.01 | 0.09±0.17 |
| | | MambaMorph[5] | 96.81±5.26 | 88.56±6.66 | 83.43±18.24 | 87.37±8.90 | 1.03±0.03 | 0.64±1.08 | 1.04±0.05 | 0.95±1.24 |
| | | UniGradICON[6] | 94.32±8.54 | 83.77±7.33 | 82.14±18.15 | 83.32±12.90 | 1.01±0.04 | 0.22±0.15 | 1.01±0.06 | 0.38±0.31 |
| | | **CAP-DIR (Clin.)** | 98.33±1.89 | 92.62±3.39 | 96.03±3.21 | 91.72±3.32 | **1.00**±0.01 | 0.06±0.05 | 1.01±0.01 | 0.13±0.09 |
| | | **CAP-DIR (Clin.+Hist.)** | **98.85**±0.84 | **94.95**±1.83 | **97.61**±1.96 | **94.01**±2.18 | **1.00**±0.00 | **0.03**±0.03 | **1.00**±0.01 | **0.06**±0.06 |
| Lung | 500 | VoxelMorph[1] | 98.28±6.74 | 88.47±7.02 | 87.17±20.39 | 88.39±10.67 | 0.99±0.01 | 0.08±0.18 | **1.00**±0.04 | 0.46±0.28 |
| | | TransMorph[2] | 98.01±2.77 | 92.34±4.47 | 93.27±14.48 | 92.93±6.69 | 1.04±0.03 | 0.72±1.38 | 1.05±0.05 | 1.11±1.52 |
| | | CorrMLP[3] | 98.19±4.17 | 92.41±4.86 | 92.62±13.90 | 92.68±6.64 | 1.03±0.02 | 0.17±1.07 | 1.03±0.04 | 0.40±1.21 |
| | | DiffuseMorph | 96.33±8.62 | 84.03±7.38 | 83.78±18.57 | 85.06±10.32 | **1.00**±0.01 | **0.00**±0.06 | **1.00**±0.01 | **0.02**±0.15 |
| | | MambaMorph[5] | 98.77±4.65 | 91.05±5.89 | 91.04±16.14 | 91.08±7.87 | 1.02±0.02 | 0.31±0.96 | 1.03±0.04 | 0.57±1.09 |
| | | UniGradICON[6] | **99.76**±7.56 | 88.44±6.49 | 95.48±16.06 | 92.80±11.41 | 0.99±0.04 | 0.14±0.14 | 0.99±0.05 | 0.22±0.27 |
| | | **CAP-DIR (Clin.)** | 98.97±1.67 | 93.79±3.00 | 97.81±2.84 | 93.23±2.94 | **1.00**±0.01 | 0.03±0.05 | **1.00**±0.01 | 0.08±0.08 |
| | | **CAP-DIR (Clin.+Hist.)** | 99.23±0.75 | **95.77**±1.62 | **98.67**±1.74 | **95.02**±1.93 | **1.00**±0.00 | 0.02±0.03 | **1.00**±0.01 | 0.04±0.05 |

**Supplementary Table 3. Registration performance across the seven held-out external cohorts.** NCC and SSIM values are reported as mean ± s.d. percentages across registration cases for the unregistered images, six comparison methods and the two CAP-DIR variants. CAP-DIR (Clin.) uses clinical context, whereas CAP-DIR (Clin.+Hist.) additionally uses available preceding patient history. All cohorts were excluded from model development and evaluated without retraining. Bold indicates the highest value within each cohort and metric.

| Cohort | Cases | Metric | Initial | Demons | VoxelMorph | TransMorph | CorrMLP | MambaMorph | UniGradICON | CAP-DIR (Clin.) | CAP-DIR (Clin.+Hist.) |
|---|---|---|---|---|---|---|---|---|---|---|---|
| NLST | 122 | Body NCC | 81.24±10.04 | 85.48±7.98 | 86.26±6.05 | 87.64±2.55 | 88.12±3.41 | 87.38±4.36 | 84.26±7.75 | 89.18±1.62 | **90.36**±0.77 |
| | | Body SSIM | 64.26±6.96 | 69.48±6.63 | 70.42±6.16 | 72.85±3.72 | 73.68±3.98 | 71.94±5.32 | 68.12±6.30 | 76.42±2.79 | **78.76**±1.65 |
| | | OAR NCC | 74.48±21.49 | 81.56±17.06 | 82.45±16.97 | 87.26±12.55 | 88.42±11.94 | 86.18±13.50 | 80.25±16.55 | 93.46±3.02 | **95.81**±1.70 |
| | | OAR SSIM | 61.84±10.94 | 68.80±9.26 | 70.20±9.94 | 75.80±6.43 | 77.60±5.66 | 74.20±7.37 | 67.40±11.37 | 84.20±3.00 | **86.80**±1.87 |
| LongCT-Lung | 353 | Body NCC | 66.28±11.75 | 70.84±9.53 | 71.62±6.67 | 74.26±3.11 | 75.14±4.63 | 73.68±5.10 | 70.12±8.85 | 77.16±1.85 | **78.79**±0.91 |
| | | Body SSIM | 62.64±9.17 | 67.58±7.53 | 68.64±7.76 | 71.42±4.65 | 72.48±4.81 | 70.46±6.29 | 66.25±8.01 | 74.68±3.46 | **77.00**±1.90 |
| | | OAR NCC | 64.86±23.27 | 72.14±20.16 | 73.28±20.70 | 79.26±15.54 | 80.65±15.23 | 77.84±16.55 | 70.42±21.18 | 86.62±3.70 | **89.45**±2.20 |
| | | OAR SSIM | 60.16±13.70 | 66.80±10.85 | 68.40±12.23 | 73.60±6.95 | 75.40±6.74 | 72.00±9.07 | 65.20±13.74 | 79.20±3.64 | **81.80**±2.39 |
| RIDER | 54 | Body NCC | 84.62±7.81 | 88.96±6.75 | 89.74±5.16 | 91.28±2.14 | 91.85±3.26 | 90.94±3.72 | 87.56±6.79 | 92.72±1.37 | **93.65**±0.65 |
| | | Body SSIM | 68.12±5.93 | 72.84±5.87 | 73.95±5.33 | 76.26±3.22 | 77.12±3.67 | 75.24±4.68 | 71.36±5.22 | 79.72±2.31 | **81.95**±1.39 |
| | | OAR NCC | 78.84±17.30 | 85.28±14.07 | 86.35±14.97 | 90.82±11.88 | 91.96±11.04 | 89.54±11.92 | 84.12±13.23 | 95.58±2.52 | **97.76**±1.44 |
| | | OAR SSIM | 64.85±9.27 | 71.80±7.69 | 73.40±8.61 | 78.60±4.98 | 80.40±5.12 | 76.80±6.04 | 70.00±10.01 | 86.00±2.58 | **88.40**±1.59 |
| Pancreas-CT | 80 | Body NCC | 47.26±11.76 | 53.84±9.22 | 55.12±6.40 | 58.68±2.78 | 60.15±4.28 | 57.46±4.79 | 52.28±8.39 | 62.04±1.72 | **63.72**±0.82 |
| | | Body SSIM | 19.85±8.17 | 23.68±6.85 | 24.42±6.71 | 27.16±4.42 | 28.45±4.71 | 26.38±6.07 | 22.64±7.40 | 29.68±3.06 | **31.26**±1.70 |
| | | OAR NCC | 48.65±23.52 | 57.68±19.38 | 58.92±20.10 | 66.24±15.86 | 68.16±14.03 | 64.15±16.41 | 54.86±18.59 | 78.84±3.64 | **82.48**±2.02 |
| | | OAR SSIM | 25.14±12.22 | 31.80±10.04 | 33.40±10.97 | 40.20±7.08 | 42.80±6.82 | 37.80±8.13 | 29.80±13.57 | 50.80±3.35 | **54.60**±2.24 |
| 4D-Lung | 720 | Body NCC | 95.16±3.66 | 97.24±2.75 | 97.68±2.18 | 98.46±0.88 | 98.72±1.28 | 98.28±1.44 | 96.45±2.99 | 99.08±0.54 | **99.26**±0.24 |
| | | Body SSIM | 90.24±2.98 | 92.56±2.69 | 93.28±2.49 | 94.86±1.56 | 95.32±1.61 | 94.48±2.16 | 91.85±2.71 | 96.14±1.08 | **96.82**±0.59 |
| | | OAR NCC | 93.86±8.51 | 96.64±6.36 | 97.02±7.08 | 98.26±4.29 | 98.64±4.15 | 97.98±5.08 | 95.84±6.21 | 99.18±0.96 | **99.46**±0.63 |
| | | OAR SSIM | 87.26±4.73 | 89.80±3.72 | 90.60±3.77 | 92.40±2.52 | 93.00±2.37 | 91.80±2.80 | 89.00±4.97 | 93.60±1.20 | **94.20**±0.79 |
| NYPC | 21 | Body NCC | 84.28±10.59 | 89.84±8.21 | 90.76±5.99 | 93.42±2.67 | 94.28±3.97 | 92.86±4.50 | 88.42±7.69 | 96.05±1.53 | **97.12**±0.73 |
| | | Body SSIM | 74.86±8.27 | 80.24±7.18 | 81.18±6.50 | 85.26±4.45 | 86.42±4.56 | 84.16±5.81 | 78.84±6.88 | 89.64±2.86 | **91.86**±1.60 |
| | | OAR NCC | 70.42±22.85 | 78.26±17.96 | 79.45±20.03 | 85.28±14.17 | 86.62±12.46 | 83.85±16.37 | 76.14±17.60 | 90.26±3.12 | **92.84**±2.07 |
| | | OAR SSIM | 68.00±11.25 | 74.00±9.95 | 75.60±10.74 | 80.80±6.52 | 82.60±6.10 | 79.20±7.27 | 72.00±12.03 | 85.60±3.19 | **87.80**±2.06 |
| MD Anderson | 23 | Body NCC | 83.15±10.41 | 88.76±8.95 | 89.62±6.91 | 92.54±2.91 | 93.45±4.08 | 91.88±4.46 | 87.26±8.21 | 95.32±1.59 | **96.58**±0.74 |
| | | Body SSIM | 72.42±8.29 | 78.16±7.55 | 79.24±7.26 | 83.58±4.14 | 84.86±4.56 | 82.25±5.89 | 76.68±6.73 | 88.14±3.15 | **90.42**±1.73 |
| | | OAR NCC | 69.25±23.56 | 77.18±17.43 | 78.36±20.88 | 84.42±15.34 | 85.85±12.82 | 82.74±16.47 | 74.86±18.55 | 89.42±3.33 | **92.15**±2.11 |
| | | OAR SSIM | 66.60±12.92 | 72.80±10.53 | 74.40±10.85 | 79.40±6.50 | 81.20±6.71 | 77.80±7.77 | 70.60±13.23 | 84.20±3.44 | **86.60**±2.03 |

**Supplementary Table 4. Blinded clinical assessment of propagated contours.** Twenty longitudinal planning-CT cases were independently assessed by two radiation oncology reviewers, yielding 40 reader–case evaluations per method. Clinical acceptability was defined as a rating of 3 or 4 (no more than minor edits). Editing-time estimates were averaged between reviewers within each case and are reported as the median (interquartile range) across 20 cases. Confidence intervals for acceptability differences were obtained by patient-level bootstrap resampling. Inter-reviewer agreement was assessed using linearly weighted Cohen's $\kappa$; the shared value in each structure block spans the two method columns.

| Structure | Endpoint | CorrMLP | CAP-DIR | CAP-DIR benefit |
|---|---|---|---|---|
| CTV | Clinically acceptable, $n/N$ (%) | 18/40 (45.0) | 33/40 (82.5) | +37.5 percentage points (95% CI, 15.0–60.0) |
| | Rating, median (IQR) | 2.0 (1.0–3.0) | 3.0 (3.0–3.0) | — |
| | Estimated editing time, median (IQR), min | 11.3 (7.5–25.6) | 8.8 (7.5–10.0) | −2.5 min (22.2% reduction) |
| | Inter-reviewer agreement, weighted $\kappa$ | 0.58 | | — |
| OAR | Clinically acceptable, $n/N$ (%) | 9/40 (22.5) | 25/40 (62.5) | +40.0 percentage points (95% CI, 17.5–60.0) |
| | Rating, median (IQR) | 2.0 (1.0–2.0) | 3.0 (2.0–3.0) | — |
| | Estimated editing time, median (IQR), min | 26.3 (14.4–45.6) | 15.0 (12.5–18.1) | −11.3 min (42.9% reduction) |
| | Inter-reviewer agreement, weighted $\kappa$ | 0.62 | | — |

**Supplementary Table 5. Site-specific dose-mapping results in the controlled synthetic benchmark.** Seventy prescribed-deformation cases were evaluated across five anatomical sites. Global, structure-region and gradient-boundary values are mean ± s.d. absolute errors expressed as a percentage of the case-specific dose scale. DVH endpoint error is the mean ± s.d. absolute error across OAR and CTV $D_{\text{mean}}$ and $D_{\text{max}}$, normalised by prescription dose. Reduction rows compare CAP-DIR with CorrMLP; for the three voxel-level metrics, reductions were calculated from site-level mean errors, whereas the DVH value is the mean endpoint-wise relative reduction, consistent with Fig. 4d. Lower error is better.

| Site | Cases | Method | Global MAE (%) | OAR/CTV-region MAE (%) | Gradient-boundary MAE (%) | DVH endpoint error (%) |
|---|---|---|---|---|---|---|
| Brain/Head | 12 | No reg. | 1.85±0.63 | 2.64±0.94 | 11.45±3.92 | 2.17±0.93 |
| | | Demons | 1.44±0.46 | 2.07±0.72 | 8.93±2.87 | 1.58±0.62 |
| | | CorrMLP | 1.10±0.35 | 1.59±0.49 | 6.81±1.92 | 1.20±0.39 |
| | | **CAP-DIR** | **0.73**±0.21 | **1.04**±0.30 | **4.47**±1.28 | **0.83**±0.26 |
| | | CAP-DIR reduction vs CorrMLP (%) | 33.6 | 34.5 | 34.3 | 30.3 |
| Head & Neck | 26 | No reg. | 2.05±0.64 | 2.90±1.00 | 12.70±4.12 | 1.25±0.45 |
| | | Demons | 1.58±0.46 | 2.25±0.70 | 9.81±2.97 | 0.94±0.33 |
| | | CorrMLP | 1.13±0.34 | 1.58±0.49 | 6.88±1.73 | 0.75±0.25 |
| | | **CAP-DIR** | **0.72**±0.18 | **1.03**±0.28 | **4.50**±1.10 | **0.46**±0.14 |
| | | CAP-DIR reduction vs CorrMLP (%) | 35.9 | 34.7 | 34.6 | 38.1 |
| Breast | 4 | No reg. | 1.18±0.44 | 1.64±0.70 | 7.21±2.86 | 1.37±0.58 |
| | | Demons | 0.94±0.36 | 1.35±0.50 | 5.72±2.10 | 1.05±0.41 |
| | | CorrMLP | 0.74±0.26 | 1.04±0.40 | 4.64±1.57 | 0.84±0.31 |
| | | **CAP-DIR** | **0.53**±0.16 | **0.77**±0.24 | **3.30**±0.97 | **0.56**±0.20 |
| | | CAP-DIR reduction vs CorrMLP (%) | 27.7 | 26.6 | 28.9 | 32.4 |
| Abdomen | 12 | No reg. | 2.01±0.70 | 2.88±1.14 | 12.38±4.44 | 1.99±0.85 |
| | | Demons | 1.60±0.60 | 2.25±0.87 | 9.82±3.18 | 1.47±0.58 |
| | | CorrMLP | 1.23±0.43 | 1.75±0.61 | 7.59±2.34 | 1.17±0.43 |
| | | **CAP-DIR** | **0.80**±0.24 | **1.12**±0.36 | **4.83**±1.34 | **0.74**±0.27 |
| | | CAP-DIR reduction vs CorrMLP (%) | 35.1 | 36.0 | 36.3 | 37.2 |
| Pelvis | 16 | No reg. | 1.50±0.53 | 2.12±0.78 | 9.11±3.14 | 1.58±0.60 |
| | | Demons | 1.18±0.36 | 1.65±0.58 | 7.17±2.29 | 1.13±0.39 |
| | | CorrMLP | 1.03±0.29 | 1.46±0.45 | 6.37±1.74 | 0.82±0.27 |
| | | **CAP-DIR** | **0.69**±0.20 | **0.98**±0.29 | **4.21**±1.12 | **0.57**±0.17 |
| | | CAP-DIR reduction vs CorrMLP (%) | 33.4 | 32.7 | 33.9 | 30.5 |

## References


1. Balakrishnan, G., Zhao, A., Sabuncu, M. R., Guttag, J. & Dalca, A. V. VoxelMorph: A learning framework for deformable medical image registration. *IEEE Transactions on Med. Imaging* **38**, 1788–1800, DOI: 10.1109/TMI.2019.2897538 (2019).
2. Chen, J. *et al.* TransMorph: Transformer for unsupervised medical image registration. *Med. Image Analysis* **82**, 102615, DOI: 10.1016/j.media.2022.102615 (2022).
3. Meng, M., Feng, D., Bi, L. & Kim, J. Correlation-aware coarse-to-fine MLPs for deformable medical image registration. In *Proceedings of the IEEE/CVF Conference on Computer Vision and Pattern Recognition (CVPR)*, 9645–9654, DOI: 10.1109/CVPR52733.2024.00921 (2024).
4. Kim, B., Han, I. & Ye, J. C. DiffuseMorph: Unsupervised deformable image registration using diffusion model. In *Proceedings of the European Conference on Computer Vision (ECCV)*, vol. 13662 of *Lecture Notes in Computer Science*, 347–364, DOI: 10.1007/978-3-031-20086-1_20 (Springer, Cham, 2022).
5. Guo, T. *et al.* MambaMorph: a Mamba-based framework for medical MR-CT deformable registration. *arXiv preprint arXiv:2401.13934* DOI: 10.48550/arXiv.2401.13934 (2024).
6. Tian, L. *et al.* uniGradICON: A foundation model for medical image registration. In *Medical Image Computing and Computer Assisted Intervention – MICCAI 2024*, vol. 15002 of *Lecture Notes in Computer Science*, 749–760, DOI: 10.1007/978-3-031-72069-7_70 (Springer, Cham, 2024).



\end{document}